\documentclass[12pt]{article}
\usepackage{makeidx}
\usepackage{amssymb}
\usepackage{amsfonts}
\usepackage{amsmath}
\usepackage{graphicx}
\usepackage{setspace}
\usepackage{authblk}
\usepackage{units}
\usepackage{appendix}
\newcommand{\lambdabar}{{\mkern0.75mu\mathchar '26\mkern -9.75mu\lambda}}
\usepackage[left=2cm,top=2cm,right=2cm,bottom=2cm]{geometry}

\begin{document}

\title{\textbf{Violation of vacuum stability by inverse square electric fields}}
\author[1,2]{T. C. Adorno\thanks{tg.adorno@gmail.com}}
\author[2,3]{S. P. Gavrilov\thanks{gavrilovsergeyp@yahoo.com, gavrilovsp@herzen.spb.ru}}
\author[2,4,5]{D. M. Gitman\thanks{gitman@if.usp.br}}
\affil[1]{\textit{College of Physical Science and Technology, Department of Physics, Hebei University, Wusidong Road 180, 071002, Baoding, China;}}
\affil[2]{\textit{Department of Physics, Tomsk State University, Lenin Prospekt 36, 634050, Tomsk, Russia;}}
\affil[3]{\textit{Department of General and Experimental Physics, Herzen State Pedagogical University of Russia, Moyka embankment 48, 191186, St. Petersburg, Russia;}}
\affil[4]{\textit{P. N. Lebedev Physical Institute, 53 Leninskiy prospekt,
119991, Moscow, Russia;}}
\affil[5]{\textit{Instituto de F\'{\i}sica, Universidade de S\~{a}o Paulo, Caixa Postal 66318, CEP 05508-090, S\~{a}o Paulo, S.P., Brazil;}}

\maketitle

\onehalfspacing

\begin{abstract}
In the framework of QED with a strong background, we study particle creation (the Schwinger effect) by a time-dependent inverse square electric field. To this end corresponding exact in- and out-solutions of the Dirac and Klein-Gordon equations are found. We calculate the vacuum-to-vacuum probability and differential and total mean numbers of pairs created from the vacuum. For electric fields varying slowly in time, we present detailed calculations of the Schwinger effect and discuss possible asymptotic regimes. The obtained results are consistent with universal estimates of the particle creation effect by electric fields in the locally constant field approximation. Differential and total quantities corresponding to asymmetrical configurations are also
discussed in detail. Finally, the inverse square electric field is used to imitate switching on and off processes. Then the case under consideration is compared with the one where an exponential electric field is used to imitate switching on and off processes.


Particle creation, Schwinger effect, time-dependent external field, Dirac and Klein-Gordon equations.
\end{abstract}

\section{Introduction\label{Sec1}}

Particle creation from the vacuum by strong external electromagnetic and
gravitational fields (sometimes we call this effect a violation of the
vacuum stability) has been studied for a long time, see, for example, Refs. ~%
\cite%
{Sch51,Nikis70a,Gitman77,FGS,GMR85,BirDav82,GriMaM,GavGit96,GavGiT08,RufVSh10,GelTan15,x-case16,slow-var17}%
. The effect can be observable if the external fields are sufficiently
strong, e.g. the magnitude of an electric field should be comparable with
the Schwinger critical field $E_{\mathrm{c}}=m^{2}c^{3}/e\hslash \simeq
10^{16}\mathrm{V/cm}$. Nevertheless, recent progress in laser physics allows
one to hope that an experimental observation of the effect can be possible
in the near future, see Refs.~\cite{Dun09} for a review. Moreover,
electron-hole pair creation from the vacuum becomes also an observable in
laboratory conditions effect in graphene and similar nanostructures, see,
e.g. Refs. \cite{dassarma}. Depending on the strong field structure,
different approaches have been proposed for calculating the effect
nonperturbatively. When a semiclassical approximation is not applicable,%
{\Large \ }the most consistent consideration is formulated in the framework
of a quantum field theory, in particular, in the framework of QED, see Refs. 
\cite{Gitman77,FGS,x-case16}. A calculation technics is based on the
existence of exact solutions of the Dirac equation with the corresponding
external field. Until now, there are known\emph{\ }only few exactly solvable
cases that allow one to apply directly such a technics. In such a way can be
calculated particle creation in the constant uniform electric field \cite%
{Sch51,Nikis70a}\emph{,} in the adiabatic electric field $E\left( t\right)
=E\cosh ^{-2}\left( t/T_{\mathrm{S}}\right) \,$\cite{NarNik70}, in the
so-called $T$-constant electric field \cite{BagGitS75,GavGit96}, in a
periodic alternating in time electric field \cite{NarNik74}, in an and
exponentially growing and decaying electric fields\ \cite%
{AdoGavGit14,AGG16,AFGG17} (see Ref. \cite{AdoGavGit17} for the
review), and in several constant inhomogeneous electric fields of similar
forms where the time $t$ is replaced by the spatial coordinate $x$. An
estimation of the role of switching on and off effects for the pair creation
effect was done in Ref. \cite{AdoFerGavGit18}.

In the present article we study the vacuum instability in an inverse square
electric field (an electric field that is inversely proportional
to time squared); see its exact definition in the next section. This
behavior is characteristic for an effective mean electric field in graphene,
which is a deformation of the initial constant electric field by
backreaction due to the vacuum instability; see Ref. \cite{GavGitY12}. From
the technical point of view, it should be noted that the problem of the
vacuum instability caused by a constant electric field in the de Sitter
space considered in Refs. \cite%
{Birrell79,Garriga94,HaoChe13,CaiKim14,Frob14,KobAsh14,StaStrXue16,TakFujYok16,ShaSin17}
shares some similarities to the above problem in the Minkowski space-time.
In addition, an inverse square electric field is useful to study the
one-loop Heisenberg-Euler effective action in the framework of a locally
constant field approximation \cite{Karb17}. At last, results of our study
allow one to better understand the role of switching on and off effects in
the violation of the vacuum stability. In Sec. \ref{Sec2} we present, for
the first time, exact solutions of the Dirac and Klein-Gordon equations with the inverse square electric field in the Minkowski space-time. With 
the help of these solutions, we study in detail the vacuum instability in
such a background in the framework of QED with $t$-electric potential steps,
using notation and some technical results of our review article \cite%
{AdoGavGit17}. In particular, differential and total mean numbers of
particles created from the vacuum are calculated in Sec. \ref{Sec3} within
the slowly varying approximation . The case of an asymmetric
configuration of the inverse square electric field is discussed in Sec. \ref%
{Sec4}. In Sec. \ref{Sec5}, the inverse square electric fields is used to
imitate switching on and off processes. The obtained results are compared
with the case when the form of switching on and off is exponential. Sec. \ref%
{Sec6} contains some concluding remarks.

\section{Solutions of wave equations with the background under consideration 
\label{Sec2}}

In this section we introduce the time dependent external electric field (in $%
D$ spatial dimensions), that switches on at the infinitely remote past $%
t=-\infty $, switches off at the infinitely remote future $t=+\infty $ and
it is inversely proportional to time squared. In what follows, we call such
a field inverse square electric field. The field is homogeneously
distributed over space, directed along the axis $x^{1}=x,$ i.e., $\mathbf{E}%
=(E\left( t\right) ,0,...,0),\ E^{i}=0$, $i=2,...,D$,%
\begin{equation}
E\left( t\right) =E\left\{ 
\begin{array}{ll}
\left( 1-t/\tau _{1}\right) ^{-2}\,, & t\in \mathrm{I}=\left( -\infty
,0\right) \,, \\ 
\left( 1+t/\tau _{2}\right) ^{-2}\,, & t\in \mathrm{II}=\left[ 0,+\infty
\right) \,.%
\end{array}%
\right.  \label{2.0}
\end{equation}%
and is specified by the potentials $A^{0}=0,\ \mathbf{A}=(A_{x}\left(
t\right) ,0,...,0)$, $\ A^{i}=0$,%
\begin{equation}
\ A_{x}\left( t\right) =E\left\{ 
\begin{array}{ll}
\tau _{1}\left[ 1-\left( 1-t/\tau _{1}\right) ^{-1}\right] \,, & t\in 
\mathrm{I}\,, \\ 
\tau _{2}\left[ \left( 1+t/\tau _{2}\right) ^{-1}-1\right] \,, & t\in 
\mathrm{II}\,.%
\end{array}%
\right.  \label{2.0.1}
\end{equation}

\begin{figure}[th!]
\begin{center}
\includegraphics[scale=0.55]{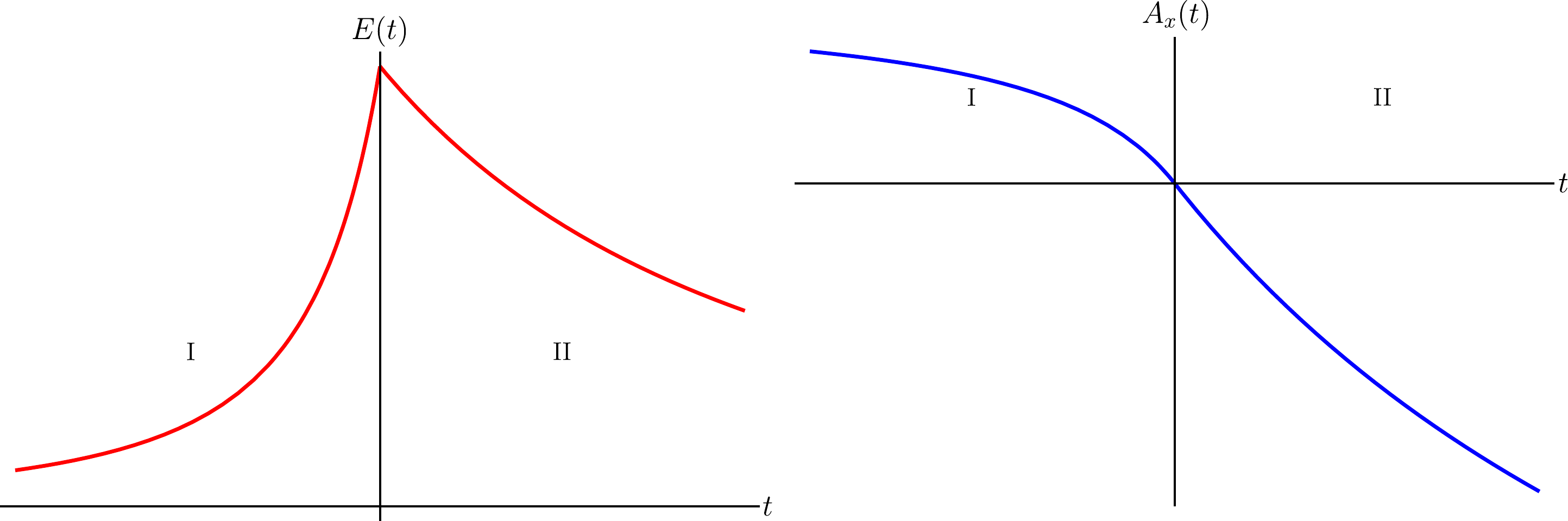}
\end{center}
\caption{(color online) The
electric field (left panel - red lines) and its potential (right panel -
blue lines) for some pulse durations $\protect\tau _{j}$ and a fixed
amplitude $E$. In both pictures, $\protect\tau _{1}<\protect\tau _{2}$.}
\label{Fig0}
\end{figure}

The inverse square electric field belongs to the so-called class of $t$%
-electric potential steps \cite{AdoGavGit17}. It is parameterized by two
constants $\tau _{1,2}$ which play the role of time scales for the pulse
durations\emph{, }respectively. The electric field (\ref{2.0}) and its
potential (\ref{2.0.1}) are pictured on Fig. \ref{Fig0} for some values of $%
\tau _{1,2}$.

For the field under consideration, Dirac spinors in a $d=D+1$ dimensional
Minkowski space-time can always be presented in the following form \cite%
{GavGit96,AdoGavGit17}\footnote{$\psi (x)$ is a $2^{[d/2]}$-component spinor
($[d/2]$ stands for the integer part of $d/2$), $m$ denotes the electron
mass and $\gamma ^{\mu }$ are Dirac matrices in $d$ dimensions. We use the
relativistic units $\hslash =c=1$, in which the fine structure constant is $%
\alpha =e^{2}/\hslash c=e^{2}$.},%
\begin{eqnarray}
&&\psi _{n}\left( x\right) =\left[ i\partial _{t}+H\left( t\right) \right]
\gamma ^{0}\exp \left( i\mathbf{pr}\right) \varphi _{n}\left( t\right)
v_{\chi ,\sigma }\,,  \nonumber \\
&&H\left( t\right) =\gamma ^{0}\left\{ \gamma ^{1}\left[ p_{x}-U\left(
t\right) \right] +\boldsymbol{\gamma }\mathbf{p}_{\perp }+m\right\} \,,
\label{2.1}
\end{eqnarray}%
where $v_{\chi ,\sigma }$ is a set of constant and orthonormalized spinors, $%
\varphi _{n}\left( t\right) $ is a scalar function, and $U\left( t\right)
=-eA\left( t\right) $ is the potential energy of an electron $\left(
e>0\right) $. The constant spinors obey the identities $\gamma ^{0}\gamma
^{1}v_{\pm ,\sigma }=\pm v_{\pm ,\sigma }$, $v_{\chi ,\sigma }^{\dag
}v_{\chi ^{\prime },\sigma ^{\prime }}=\delta _{\chi ,\chi ^{\prime }}\delta
_{\sigma ,\sigma ^{\prime }}$, in which $\sigma =\left\{ \sigma _{1},\sigma
_{2},\dots ,\sigma _{\lbrack d/2]-1}\right\} $ represents a set of
eigenvalues of additional spin operators compatible with $\gamma ^{0}\gamma
^{1}$, while the scalar function $\varphi _{n}\left( t\right) $ satisfy the
second-order ordinary differential equation%
\begin{equation}
\left\{ \frac{d^{2}}{dt^{2}}+\left[ p_{x}-U\left( t\right) \right] ^{2}+\pi
_{\perp }^{2}-i\chi \dot{U}\left( t\right) \right\} \varphi _{n}\left(
t\right) =0\,,\ \ \pi _{\perp }=\sqrt{\mathbf{p}_{\perp }^{2}+m^{2}}\,.
\label{s2}
\end{equation}

Introducing new variables,%
\begin{eqnarray}
z_{1}\left( t\right) &=&2i\omega _{1}\tau _{1}\left( 1-t/\tau _{1}\right)
\,,\ \ t\in \mathrm{I}\,,  \nonumber \\
z_{2}\left( t\right) &=&2i\omega _{2}\tau _{2}\left( 1+t/\tau _{2}\right)
\,,\ \ t\in \mathrm{II}\,,  \label{21.1}
\end{eqnarray}%
one can reduce Eq. (\ref{s2}) to the Whittaker differential equation%
\footnote{%
Hereafter, the index $j=\left( 1,2\right) $ distinguish quantities
associated with the first interval \textrm{I }$\left( j=1\right) $ from the
second interval \textrm{II} $\left( j=2\right) $.} \cite%
{Whittaker,WhiWat,DLMF}%
\begin{equation}
\left( \frac{d^{2}}{dz_{j}^{2}}-\frac{1}{4}+\frac{\kappa _{j}}{z_{j}}+\frac{%
1/4-\mu _{j}^{2}}{z_{j}^{2}}\right) \varphi _{n}\left( t\right) =0\,.
\label{21.3}
\end{equation}%
where%
\begin{eqnarray}
&&\kappa _{j}=-\left( -1\right) ^{j}ieE\tau _{j}^{2}\pi _{j}/\omega _{j}\,,\
\ \mu _{j}=\left( -1\right) ^{j}\left( ieE\tau _{j}^{2}+\chi /2\right) \,, 
\nonumber \\
&&\omega _{j}=\sqrt{\pi _{j}^{2}+\pi _{\perp }^{2}}\,,\ \ \pi
_{j}=p_{x}-\left( -1\right) ^{j}eE\tau _{j}\,.  \label{21.4}
\end{eqnarray}%
A fundamental set of solutions of Eq. (\ref{21.3}) can then be represented
as a linear combination of Whittaker functions,%
\begin{eqnarray}
&&\varphi _{n}\left( t\right) =b_{1}^{j}W_{\kappa _{j},\mu _{j}}\left(
z_{j}\right) +b_{2}^{j}W_{-\kappa _{j},\mu _{j}}\left( e^{-i\pi
}z_{j}\right) \,,  \nonumber \\
&&W_{\kappa _{j},\mu _{j}}\left( z_{j}\right)
=e^{-z_{j}/2}z_{j}^{c_{j}/2}\Psi \left( a_{j},c_{j};z_{j}\right) \,, 
\nonumber \\
&&W_{-\kappa _{j},\mu _{j}}\left( e^{-i\pi }z_{j}\right) =e^{-i\pi
c_{j}/2}e^{z_{j}/2}z_{j}^{c_{j}/2}\Psi \left( c_{j}-a_{j},c_{j};e^{-i\pi
}z_{j}\right) \,,  \label{21.5}
\end{eqnarray}%
where $a_{j}=\mu _{j}-\kappa _{j}+1/2$, $c_{j}=1+2\mu _{j},$ $b_{1,2}^{j}$
are some arbitrary constants, and $\Psi \left( a,c;z\right) $ are confluent
hypergeometric functions (CHFs) \cite{Erdelyi}.

By definition, the electric field (\ref{2.0}) vanishes at the infinitely
remote past $\left( t=-\infty \right) $ and at the infinitely remote future $%
\left( t=+\infty \right) $, which means that particles must be free at these
limits. From the asymptotic properties of Whittaker functions with large
argument\footnote{%
Originally, Whittaker \cite{Whittaker,WhiWat} wrote this asymptotic form for
a different domain in the $z$-complex plane, namely $\left\vert \arg
z\right\vert \leq \pi -0^{+}$, by expanding the binomial inside of his
integral representation for $W_{\kappa ,\mu }\left( z\right) $. However, as
discussed in \cite{Buchholz}, the domain changes to $\left\vert \arg \left(
z\right) \right\vert \leq 3\pi /2-0^{+}$ by rotating the path of integration
over an angle near $\pi /2$ in any direction.} \cite{DLMF},%
\begin{equation}
W_{\kappa ,\mu }\left( z\right) =e^{-z/2}z^{\kappa }\left[ 1+O\left(
z^{-1}\right) \right] \,,\ \ z\rightarrow \infty \,,\ \ \left\vert \arg
z\right\vert \leq 3\pi /2-0^{+}\,,  \label{21.8}
\end{equation}%
one may classify exact solutions for first $\mathrm{I}$ and second \textrm{II%
} intervals according to their asymptotic behavior as free particles $%
\left\{ \ _{+}\varphi \left( t\right) ,\ ^{+}\varphi \left( t\right)
\right\} $ or free antiparticles $\left\{ \ _{-}\varphi \left( t\right) ,\
^{-}\varphi \left( t\right) \right\} $ as follows:%
\begin{eqnarray}
\ _{+}\varphi _{n}\left( t\right) &=&\,_{+}\mathcal{N}W_{-\kappa _{1},\mu
_{1}}\left( e^{-i\pi }z_{1}\right) ,\ _{-}\varphi _{n}\left( z_{1}\right)
=\,_{-}\mathcal{N}W_{\kappa _{1},\mu _{1}}\left( z_{1}\right) \,,\ \ t\in 
\mathrm{I}\,,  \nonumber \\
\ ^{-}\varphi _{n}\left( t\right) &=&\,^{-}\mathcal{N}W_{-\kappa _{2},\mu
_{2}}\left( e^{-i\pi }z_{2}\right) ,\ ^{+}\varphi _{n}\left( z_{2}\right)
=\,^{+}\mathcal{N}W_{\kappa _{2},\mu _{2}}\left( z_{2}\right) \,,\ \ t\in 
\mathrm{II}\,,  \label{21.10}
\end{eqnarray}%
Here, the constants$\ _{\pm }\mathcal{N},\ ^{\pm }\mathcal{N}$ are
conveniently chosen in order to normalize Dirac spinors with respect to the
equal-time inner product $\left( \psi ,\psi ^{\prime }\right) =\int d\mathbf{%
x}\psi ^{\dagger }\left( x\right) \psi ^{\prime }\left( x\right) $. After
the usual volume regularization, we obtain%
\[
\left\vert \ _{\pm }\mathcal{N}\right\vert =\frac{\exp \left( -i\pi \kappa
_{1}/2\right) }{\sqrt{2\omega _{1}V_{\left( d-1\right) }q_{1}^{\mp \chi }}}%
\,,\ \ \left\vert \ ^{\pm }\mathcal{N}\right\vert =\frac{\exp \left( -i\pi
\kappa _{2}/2\right) }{\sqrt{2\omega _{2}V_{\left( d-1\right) }q_{2}^{\mp
\chi }}}\,, 
\]%
where $q_{j}^{\mp \chi }=\omega _{j}\mp \chi \pi _{j}$ and $V_{\left(
d-1\right) }$ is the volume of the $D$-dimensional Euclidean space.

With the help of Eq. (\ref{21.10}), we use Eq. (\ref{2.1}) to introduce IN $%
\left\{ \ _{\zeta }\psi \left( x\right) \right\} $ and OUT $\left\{ \
^{\zeta }\psi \left( x\right) \right\} $ sets of solutions of Dirac equation
corresponding to free electrons $\left( \zeta =+\right) $ or free positrons $%
\left( \zeta =-\right) $ at $t\rightarrow \pm \infty $. Both sets are
related \textit{via} linear transformations, for instance$\ ^{\zeta }\psi
_{n}\left( x\right) =\sum_{\zeta ^{\prime }}g\left( _{\zeta ^{\prime
}}|^{\zeta }\right) \ _{\zeta ^{\prime }}\psi _{n}\left( x\right) $, where
coefficients $g\left( _{\zeta }|^{\zeta ^{\prime }}\right) $ are diagonal $%
\left( \ _{\zeta }\psi _{n},\ ^{\zeta ^{\prime }}\psi _{n^{\prime }}\right)
=g\left( _{\zeta }|^{\zeta ^{\prime }}\right) \delta _{nn^{\prime }}$ and
obey the properties%
\begin{equation}
g\left( ^{\zeta ^{\prime }}|_{\zeta }\right) ^{\ast }=g\left( _{\zeta
}|^{\zeta ^{\prime }}\right) \,,\ \sum_{\zeta ^{\prime }}g\left( ^{\zeta
}|_{\zeta ^{\prime }}\right) g\left( _{\zeta ^{\prime }}|^{\zeta ^{\prime
\prime }}\right) =\delta _{\zeta ,\zeta ^{\prime \prime }}\,,  \label{21.14}
\end{equation}%
This implies decompositions for scalar functions as follows\footnote{%
We conveniently introduce an auxiliary constant $\kappa $ to extend results
to scalar QED, in which $\kappa =-1$. It should not be confused with the
parameters of the Whittaker functions $\kappa _{j}$, defined by Eq. (\ref%
{21.4}).}%
\begin{eqnarray*}
\ ^{+}\varphi _{n}\left( t\right) &=&g\left( _{+}|^{+}\right) \ _{+}\varphi
_{n}\left( t\right) +\kappa g\left( _{-}|^{+}\right) \ _{-}\varphi
_{n}\left( t\right) \,, \\
\ _{-}\varphi _{n}\left( t\right) &=&g\left( ^{+}|_{-}\right) \ ^{+}\varphi
_{n}\left( t\right) +\kappa g\left( ^{-}|_{-}\right) \ ^{-}\varphi
_{n}\left( t\right) \,.
\end{eqnarray*}%
Using these decompositions and continuity conditions%
\[
\left. \ _{-}^{+}\varphi _{n}\left( t\right) \right\vert _{t=0+\varepsilon
}=\left. \ _{-}^{+}\varphi _{n}\left( t\right) \right\vert _{t=0-\varepsilon
}\,,\ \ \left. \partial _{t}\ _{-}^{+}\varphi _{n}\left( t\right)
\right\vert _{t=0-\varepsilon }=\left. \partial _{t}\ _{-}^{+}\varphi
_{n}\left( t\right) \right\vert _{t=0+\varepsilon }\,, 
\]%
one can calculate basic coefficients,%
\begin{eqnarray}
&&g\left( _{-}|^{+}\right) =2\kappa e^{\frac{i\pi \chi }{2}}e^{i\theta _{+}}%
\sqrt{\frac{\tau _{1}q_{1}^{+\chi }\tau _{2}}{q_{2}^{-\chi }}}\left( \frac{%
\omega _{2}\tau _{2}}{\omega _{1}\tau _{1}}\right) ^{\frac{\chi }{2}}e^{-%
\frac{\pi }{2}\left( \nu _{1}^{-}+\nu _{2}^{+}\right) }\Delta \left( t\right)
\nonumber \\
&&\Delta \left( t\right) =\Psi \left( a_{2},c_{2};z_{2}\right)
f_{1}^{+}\left( t\right) +\Psi \left( c_{1}-a_{1},c_{1};e^{-i\pi
}z_{1}\right) f_{2}^{-}\left( t\right) \,;  \label{21.15} \\
&&g\left( ^{+}|_{-}\right) =2e^{-\frac{i\pi \chi }{2}}e^{i\theta _{-}}\sqrt{%
\frac{\tau _{1}q_{2}^{-\chi }\tau _{2}}{q_{1}^{+\chi }}}\left( \frac{\omega
_{2}\tau _{2}}{\omega _{1}\tau _{1}}\right) ^{\frac{\chi }{2}}e^{\frac{\pi }{%
2}\left( \nu _{1}^{+}+\nu _{2}^{-}\right) }\tilde{\Delta}\left( t\right) \,,
\nonumber \\
&&\tilde{\Delta}\left( t\right) =\Psi \left( a_{1},c_{1};z_{1}\right)
f_{2}^{+}\left( t\right) +\Psi \left( c_{2}-a_{2},c_{2};e^{-i\pi
}z_{2}\right) f_{1}^{-}\left( t\right) \,.  \label{21.16}
\end{eqnarray}%
Here $\kappa =+1$,%
\[
\nu _{j}^{\pm }=eE\tau _{j}^{2}\left( 1\pm \frac{\pi _{j}}{\omega _{j}}%
\right) \,,\ \ \theta _{\pm }=\pm \left( \omega _{1}\tau _{1}-\omega
_{2}\tau _{2}\right) -eE\left[ \tau _{1}^{2}\ln \left( 2\omega _{1}\tau
_{1}\right) -\tau _{2}^{2}\ln \left( 2\omega _{2}\tau _{2}\right) \right]
\,, 
\]%
and $f_{j}^{\pm }\left( t\right) $ are combinations of CHFs and their
derivatives%
\begin{eqnarray}
&&f_{j}^{+}\left( t\right) =\omega _{j}\left[ \frac{1}{2}\left( 1+\frac{c_{j}%
}{z_{j}}\right) +\frac{d}{dz_{j}}\right] \Psi \left(
c_{j}-a_{j},c_{j};e^{-i\pi }z_{j}\right) \,,  \nonumber \\
&&f_{j}^{-}\left( t\right) =\omega _{j}\left[ \frac{1}{2}\left( -1+\frac{%
c_{j}}{z_{j}}\right) +\frac{d}{dz_{j}}\right] \Psi \left(
a_{j},c_{j};z_{j}\right) \,.  \label{21.17}
\end{eqnarray}

It can be seen that the calculated coefficients can be mapped onto one
another through the simultaneous exchanges $p_{x}\rightleftarrows -p_{x}$
and $\tau _{1}\rightleftarrows \tau _{2}$. For example, taking into account
that $\Psi \left( a_{2},c_{2};z_{2}\right) \rightleftarrows \Psi \left(
a_{1}-c_{1}+1,2-c_{1};z_{1}\right) $ and $\Psi \left(
c_{1}-a_{1},c_{1};e^{-i\pi }z_{1}\right) \rightleftarrows \Psi \left(
1-a_{2},2-c_{2};e^{-i\pi }z_{2}\right) $ under these exchanges and using
some Kummer transformations (see e.g. \cite{Erdelyi}),%
\begin{eqnarray*}
&&\Psi \left( a_{1}-c_{1}+1,2-c_{1};z_{1}\right) =z_{1}^{c_{1}-1}\Psi \left(
a_{1},c_{1};z_{1}\right) \,, \\
&&\Psi \left( 1-a_{2},2-c_{2};e^{-i\pi }z_{2}\right) =e^{i\pi \left(
1-c_{2}\right) }z_{2}^{c_{2}-1}\Psi \left( c_{2}-a_{2},c_{2};e^{-i\pi
}z_{2}\right) \,,
\end{eqnarray*}%
one finds that $\Delta (t)\rightleftarrows e^{i\pi \left( 1-c_{2}\right)
}z_{1}^{c_{1}-1}z_{2}^{c_{2}-1}\tilde{\Delta}(t)$. The latter properties
yield the identity%
\begin{equation}
g\left( _{-}|^{+}\right) \rightleftarrows \kappa g\left( ^{+}|_{-}\right) \,,
\label{21.20}
\end{equation}%
that shall be useful in the calculation of differential quantities, as
discussed below.

\section{Quantities characterizing the vacuum instability\label{Sec3}}

The $g$'s coefficients allow us to find differential mean numbers $N_{n}^{%
\mathrm{cr}}$ of pairs created from the vacuum, the total number $N$ and the
vacuum-to-vacuum transition probability $P_{v}$:%
\begin{eqnarray}
&&N_{n}^{\mathrm{cr}}=\left\vert g\left( _{-}|^{+}\right) \right\vert
^{2}\,,\ \ N^{\mathrm{cr}}=\sum_{n}N_{n}^{\mathrm{cr}}\,,  \label{22.1} \\
&&P_{v}=\exp \left[ \kappa \sum_{n}\ln \left( 1-\kappa N_{n}^{\mathrm{cr}%
}\right) \right] \,.  \label{22.2}
\end{eqnarray}%
Once the mean numbers $N_{n}^{\mathrm{cr}}$ depends on the coefficients
given by Eqs. (\ref{21.15}) and (\ref{21.16}), its calculation can be
simplified through the properties given by Eqs. (\ref{21.14}) and (\ref%
{21.20}). For example, with $N_{n}^{\mathrm{cr}}$ calculated for $p_{x}$
negative, the corresponding expression for $p_{x}$ positive can be extracted
from these results through simple exchanges $-p_{x}\rightleftarrows p_{x}$
and $\tau _{1}\rightleftarrows \tau _{2}$. Moreover, note that all results
above can be generalized to discuss creation of Klein-Gordon particles from
the vacuum. To do so, one has to take into account that $n=\mathbf{p}$ and
substitute $\kappa =-1$, $\chi =0$, $q_{j}^{\mp \chi }=1$ in all formulas
throughout in this article.

\subsection{Slowly varying field regime\label{Sec3.1}}

\subsubsection{Differential mean numbers\label{Sec3.1.1}}

In this subsection we calculate differential mean numbers of pairs created
from the vacuum $N_{n}^{\mathrm{cr}}$ in the most favorable configuration
for particle creation, that is when the external field is sufficiently
strong and acts over a sufficiently large time. We call such configuration
as slowly varying field, which specified by the following condition%
\begin{equation}
\min \left( eE\tau _{1}^{2},eE\tau _{2}^{2}\right) \gg \max \left( 1,\frac{%
m^{2}}{eE}\right) \,,  \label{22.3}
\end{equation}%
with $\tau _{1}/\tau _{2}$ fixed. Within this condition, it is still
necessary to compare parameters involving the quantum numbers with the
numbers above. To this end, it is meaningful to discuss some general
peculiarities underlying the momentum distribution of pairs created by $t$%
-electric steps. First, since the electric field is homogeneously directed
along the $x$-direction only, it creates pairs with a wider range of values
of $p_{x}$ instead $\mathbf{p}_{\perp }$, once they are accelerated along
the direction of the field. Accordingly, one may consider a restricted range
of values to $\mathbf{p}_{\perp }$, namely $\sqrt{\lambda }<K_{\perp }$, in
which $K_{\perp }$ is any number within the interval $\min \left( eE\tau
_{1}^{2},eE\tau _{2}^{2}\right) \gg K_{\perp }^{2}\gg \max \left(
1,m^{2}/eE\right) $. As for the longitudinal momentum $p_{x}$, we restrict
subsequent considerations to $p_{x}$ negative and generalize results for $%
p_{x}$ positive using the properties discussed at the end of Sec. \ref{Sec2}%
. Thus, as $p_{x}$ admits values within the half-infinite interval $-\infty
<p_{x}\leq 0$, the kinetic momentum $\pi _{1}$ varies from large and
positive to large and negative values $eE\tau _{1}\geq \pi _{1}>-\infty $.
However, differential mean numbers $N_{n}^{\mathrm{cr}}$ are significant
only in the range $-\left\vert \pi _{\perp }\right\vert \beta _{1}\leq \pi
_{1}\leq eE\tau _{1}$, whose main contributions lies in four specific
subranges%
\begin{eqnarray}
&&\left( a\right) \ \sqrt{eE}\tau _{1}-\frac{\delta _{1}}{\sqrt{2}}\leq 
\frac{\pi _{1}}{\sqrt{eE}}\leq \sqrt{eE}\tau _{1}\,,  \nonumber \\
&&\left( b\right) \ \sqrt{eE}\tau _{1}\left( 1-\Upsilon _{1}\right) <\frac{%
\pi _{1}}{\sqrt{eE}}<\sqrt{eE}\tau _{1}-\frac{\delta _{1}}{\sqrt{2}}\,, 
\nonumber \\
&&\left( c\right) \ \sqrt{\lambda }\beta _{1}\leq \frac{\pi _{1}}{\sqrt{eE}}%
\leq \sqrt{eE}\tau _{1}\left( 1-\Upsilon _{1}\right) \,,\ \ \left( d\right)
\ \frac{\left\vert \pi _{1}\right\vert }{\sqrt{eE}}<\sqrt{\lambda }\beta
_{1}\,,  \label{22.5}
\end{eqnarray}%
wherein $0<\delta _{1}\ll 1$, $0<\beta _{1}\ll 1$ and $\delta _{1}/\sqrt{2}%
<\Upsilon _{1}\ll 1$ are sufficiently small numbers so that $\Upsilon _{1}%
\sqrt{eE}\tau _{1}$ and $\beta _{1}eE\tau _{1}^{2}$ are finite. To study the
mean numbers $N_{n}^{\mathrm{cr}}$, we conveniently introduce two sets of
variables%
\begin{equation}
\eta _{1}=\frac{e^{-i\pi }z_{1}}{c_{1}}\,,\ \ \eta _{2}=\frac{z_{2}}{c_{2}}%
\,,\ \ \mathcal{Z}_{j}=\left( \eta _{j}-1\right) \mathcal{W}_{j}\sqrt{c_{j}}%
\,,  \label{22.6}
\end{equation}%
where $\mathcal{W}_{j}=\left\vert \eta _{j}-1\right\vert ^{-1}\sqrt{2\left(
\eta _{j}-1-\ln \eta _{j}\right) }$, and take into account that $\pi _{2}$
is large and negative $\pi _{2}\leq -eE\tau _{2}$, which means that $a_{2}$
is fixed while $c_{2}$ and $z_{2}$ are large throughout the ranges above.

The range $\left( a\right) $ correspond to small values to $\left\vert
p_{x}\right\vert /\sqrt{eE}$ and values for $\eta _{1}$ and $\eta _{2}$
close to the unity,%
\begin{equation}
\left( a\right) \ 1>\eta _{1}\geq 1-\frac{\delta _{1}}{\sqrt{2eE}\tau _{1}}%
\,,\ \ 1<\eta _{2}\leq 1+\frac{\delta _{1}}{\sqrt{2eE}\tau _{2}}\,,
\label{22.7}
\end{equation}%
so that $\mathcal{Z}_{1}$ and $\mathcal{Z}_{2}$ are small in this range, $%
\left\vert \mathcal{Z}_{j}\right\vert <\delta _{1}$. As a result, one can
use Eq. (\ref{ap6}) in Appendix \ref{App1} and the approximations $\nu
_{1}^{-}=\lambda /2\left[ 1+O\left( \left( eE\tau _{1}^{2}\right)
^{-1/2}\right) \right] $, $\nu _{2}^{+}=\lambda /2\left[ 1+O\left( \left(
eE\tau _{2}^{2}\right) ^{-1/2}\right) \right] $ to show that the mean number
of pairs created (\ref{22.1}) reads%
\begin{equation}
\left( a\right) \ N_{n}^{\mathrm{cr}}\approx e^{-\pi \lambda }\,,
\label{22.8}
\end{equation}%
in leading-order approximation\footnote{%
Here and in what follows, we use the symbol \textquotedblleft $\approx $%
\textquotedblright\ to denote an asymptotic relation truncated in
leading-order approximation, under the understanding that the condition (\ref%
{22.3}) is satisfied.}. This result coincides with differential number of
created particles in a constant electric field \cite{Nikis70a}.

In the range $\left( c\right) $, $\left\vert p_{x}\right\vert /\sqrt{eE}$ is
finite $\min \left( p_{x}/\sqrt{eE}\right) =-\Upsilon _{1}\sqrt{eE}\tau _{1}$%
, the variables $\eta _{j}$ are approximately given by $\eta _{1}\sim
1-\Upsilon _{1}$, $\eta _{2}\sim 1+\Upsilon _{1}\tau _{1}/\tau _{2}$ and $%
\mathcal{Z}_{j}$ are considered large. Thus, one may use the asymptotic
approximation given by the second line of Eq. (\ref{ap10}) for $\Psi \left(
c_{1}-a_{1},c_{1};e^{-i\pi }z_{1}\right) $ and Eq. (\ref{ap8}) for $\Psi
\left( a_{2},c_{2};z_{2}\right) $, both in Appendix \ref{App1}, to obtain%
\begin{equation}
\left( c\right) \ N_{n}^{\mathrm{cr}}\approx \exp \left( -2\pi \nu
_{1}^{-}\right) \,.  \label{22.9}
\end{equation}%
Note that this distribution tends to the uniform distribution (\ref{22.8})
as $\pi _{1}\rightarrow eE\tau _{1}\left( 1-\Upsilon _{1}\right) $. Eqs. (%
\ref{22.8}) and (\ref{22.9}) are valid both for Fermions as for Bosons.

In the range $\left( d\right) $, $\left\vert p_{x}\right\vert /\sqrt{eE}$ is
large and $\eta _{2}$ is approximately given by $\eta _{2}\sim 1+\tau
_{1}/\tau _{2}$, so that $\mathcal{Z}_{2}$ is large in this interval.
Therefore, one may use the same asymptotic approximation for $\Psi \left(
a_{2},c_{2};z_{2}\right) $ as in the range $\left( c\right) $. As for the
Kummer function $\Psi \left( c_{1}-a_{1},c_{1};e^{-i\pi }z_{1}\right) $, it
is more convenient to rewrite it in terms of the Whittaker function $%
W_{-\kappa _{1},\mu _{1}}\left( e^{-i\pi }z_{1}\right) $ through the
relation (\ref{21.5}) and use the fact that $z_{1}\left( 0\right) $ and $%
\kappa _{1}$ are fixed in this interval, namely $\min z_{1}\left( 0\right)
=2i\sqrt{\lambda }\sqrt{eE}\tau _{1}$ and $\left\vert \kappa _{1}\right\vert
\leq \beta _{1}eE\tau _{1}^{2}$. As a result, one may use Eq. (\ref{ap2.3})
in \ref{App1} to show that the mean number of pairs created acquires the form%
\begin{equation}
\left( d\right) \ N_{n}^{\mathrm{cr}}\approx \frac{\exp \left( -\pi \nu
_{1}^{-}\right) }{\sinh \left( 2\pi eE\tau _{1}^{2}\right) }\times \left\{ 
\begin{array}{l}
\sinh \left( \pi \nu _{1}^{+}\right) \,,\ \ \mathrm{Fermi} \\ 
\cosh \left( \pi \nu _{1}^{+}\right) \,,\ \ \mathrm{Bose}%
\end{array}%
\right. \,.  \label{22.11}
\end{equation}%
Once the longitudinal kinetic momentum $\pi _{1}$ is small in this interval
and the conditions (\ref{22.3}) are satisfied, one may simplify the
hyperbolic functions above to obtain $N_{n}^{\mathrm{cr}}\approx \exp \left(
-2\pi \nu _{1}^{-}\right) $ in leading-order approximation. This result
agrees with the approximation obtained for the interval $\left( c\right) $
so that Eq. (\ref{22.9}) is uniform over the intervals $\left( c\right) $
and $\left( d\right) $. In the intermediate interval $\left( b\right) $, the
differential mean numbers $N_{n}^{\mathrm{cr}}$ varies between the
approximations (\ref{22.8}) and (\ref{22.9}). At this interval, the
Whittaker function $W_{-\kappa _{1},\mu _{1}}\left( e^{-i\pi }z_{1}\right) $
(or $\Psi \left( c_{1}-a_{1},c_{1};e^{-i\pi }z_{1}\right) $) has to be
considered exactly while $\Psi \left( a_{2},c_{2};z_{2}\right) $ may be
approximated by Eq. (\ref{ap1}).

Repeating the same considerations above and using the properties of the
differential mean numbers $N_{n}^{\mathrm{cr}}$ under the exchanges $%
p_{x}\rightleftarrows -p_{x}$ and $\tau _{1}\rightleftarrows \tau _{2}$
discussed in the previous section, one may easily generalize results for $%
p_{x}$ positive, $0\leq p_{x}<+\infty $. As a result, the mean numbers $%
N_{n}^{\mathrm{cr}}$ can be approximated by the asymptotic forms%
\begin{equation}
N_{n}^{\mathrm{cr}}\approx \left\{ 
\begin{array}{ll}
\exp \left( -2\pi \nu _{1}^{-}\right) & \mathrm{if\;}-\infty <p_{x}\leq 0\,,
\\ 
\exp \left( -2\pi \nu _{2}^{+}\right) & \mathrm{if\;}0<p_{x}<+\infty \,.%
\end{array}%
\right.  \label{22.13}
\end{equation}%
According to the results above, dominant contributions (\ref{22.13}) are
formed in ranges of large longitudinal kinetic momenta, namely, $\pi _{\bot
}<\pi _{1}\leqslant eE\tau _{1}$\ for $p_{x}<0$\ and as $-eE\tau _{2}<\pi
_{2}<-\pi _{\bot }$\ for $p_{x}>0$.

To extend the analysis above to a wider range of values to the longitudinal
momentum $p_{x}$ and compare asymptotic approximations with exact results,
it is useful to represent the mean numbers graphically. Thus, in Figs. \ref%
{Fig1}, \ref{Fig2}, we present the differential mean numbers of pairs
created from the vacuum $N_{n}^{\mathrm{cr}}$ given by Eq. (\ref{22.1}) as a
function of the longitudinal momentum $p_{x}$ for some values of the pulses
duration $\tau _{j}$ and amplitude $E$ equal to the critical Schwinger value 
$E=E_{\mathrm{c}}=m^{2}/e$. In addition, we include the approximations given
by Eq. (\ref{22.13}) for the same values to the pulses durations $\tau _{j}$
and amplitude $E$. In these plots, we set $\mathbf{p}_{\perp }=0$ and select
for convenience a system of units, in which $\hslash =c=m=1$. In this
system, the reduced Compton wavelength $\lambdabar _{\mathrm{c}}=\hslash
/mc=1$\ is one unit of length, the Compton time $\lambdabar _{\mathrm{c}%
}/c=\hslash /mc^{2}=1$ one unit of time and electron's rest energy $mc^{2}=1$
one unit of energy. In the plots below, the pulse durations $\tau _{j}$ and
the quantum numbers $p_{x}$ are dimensionless quantities, relative to
electron's rest mass $p_{x}/m$ and $m\tau _{j}$.

\begin{figure}[th!]
\begin{center}
\includegraphics[scale=0.48]{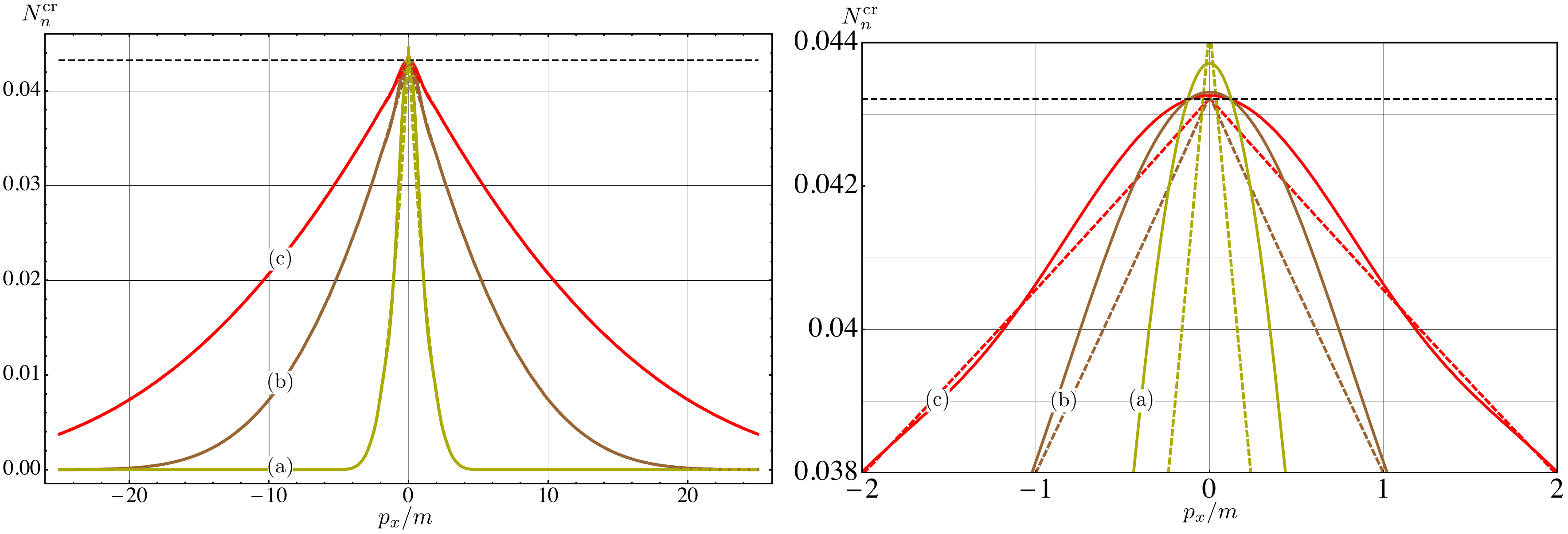}
\end{center}
\caption{(Color online) Differential mean number $N_{n}^{\mathrm{%
cr}}$ of Fermions created by the symmetric inverse square electric field (%
\protect\ref{2.0}), in which $\protect\tau _{1}=\protect\tau _{2}$. The
exact differential mean numbers (\protect\ref{22.1}) are represented by red,
brown and yellow solid lines while the asymptotic approximations (\protect
\ref{22.9}), (\protect\ref{22.13}) are represented by dashed color lines.
The right panel shows the range of larger discrepancy between exact and
asymptotic expressions. All lines labelled with $(\mathrm{a})$,\ $\left( 
\mathrm{b}\right) $ and $\left( \mathrm{c}\right) $, refers to $m\protect%
\tau =10$, $50$ and $100$, respectively. In both plots, $E=E_{\mathrm{c}}$
and the horizontal dashed line denotes the uniform distribution $e^{-\protect%
\pi \protect\lambda }$.}
\label{Fig1}
\end{figure}

\begin{figure}[th!]
\begin{center}
\includegraphics[scale=0.48]{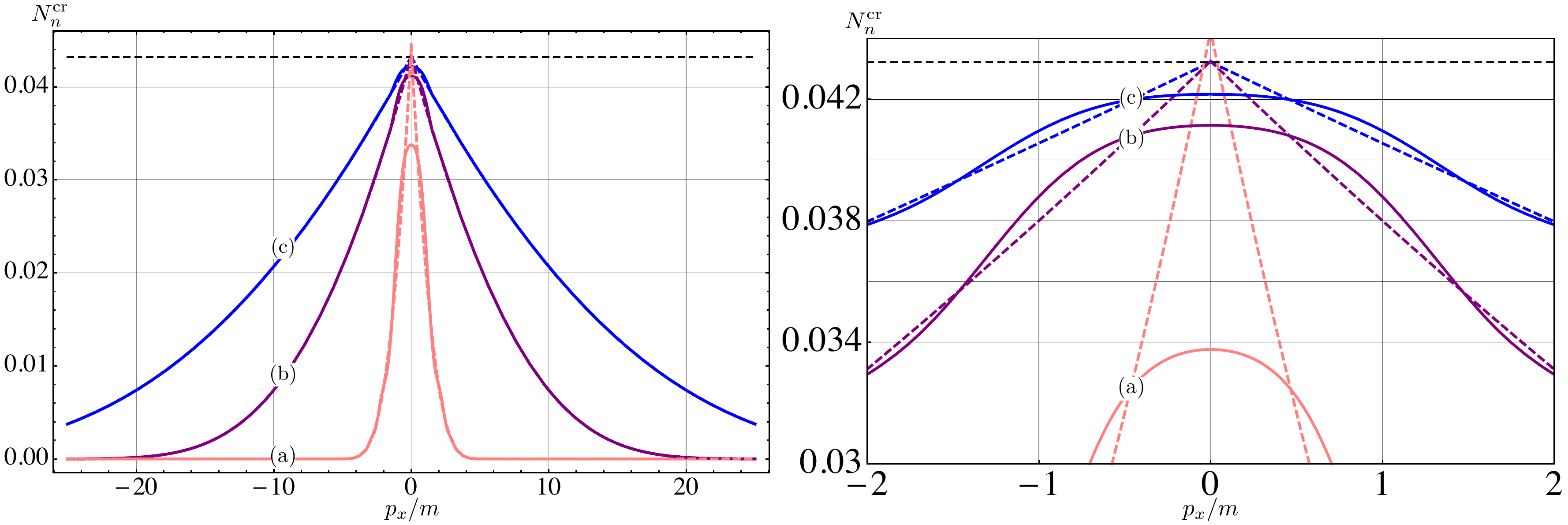}
\end{center}
\caption{(Color online)
Differential mean number $N_{n}^{\mathrm{cr}}$ of Bosons created by the
symmetric inverse square electric field (\protect\ref{2.0}), in which $%
\protect\tau _{1}=\protect\tau _{2}$. The exact differential mean numbers (%
\protect\ref{22.1}) are represented by blue, purple and pink solid lines
while the asymptotic approximations (\protect\ref{22.9}), (\protect\ref%
{22.13}) are represented by dashed color lines. The right panel shows the
range of larger discrepancy between exact and asymptotic expressions. All
lines labelled with $(\mathrm{a})$,\ $\left( \mathrm{b}\right) $ and $\left( 
\mathrm{c}\right) $, refers to $m\protect\tau =10$, $50$ and $100$,
respectively. In both plots, $E=E_{\mathrm{c}}$ and the horizontal dashed
line denotes the uniform distribution $e^{-\protect\pi \protect\lambda }$.}
\label{Fig2}
\end{figure}

According to the above results, the mean number of pairs created $N_{n}^{%
\mathrm{cr}}$ tend to the uniform distribution $e^{-\pi \lambda }$ as the
pulses duration $\tau _{j}$ increases. This is consistent with the fact that
the inverse square electric field (\ref{2.0}) tends to a constant electric
field (or a $T$-constant field with $T$ sufficiently large) as the pulses
duration $\tau _{j}$ increases, whose mean numbers are uniform over a
sufficiently wide range of values to the longitudinal momentum $p_{x}$.
Therefore, the exact distributions (\ref{22.1}) are expected approach to the
uniform distribution for sufficiently large values of the pulses duration $%
\tau _{j}$. Moreover, it is seen that the exact distributions tends to the
uniform distribution for sufficiently small values to the longitudinal
momentum $p_{x}$. This is also in agreement with the asymptotic estimate
given by Eq. (\ref{22.8}), obtained for $p_{x}$ sufficiently small. Finally,
comparing asymptotic approximations (dashed lines) with exact distributions
(solid lines), we conclude that the accuracy of the approximations (\ref%
{22.13}) increase as $m\tau $ increases. This results from the fact that as $%
m\tau $ increases, the parameter $eE\tau ^{2}$ increases as well. Thus,
larger values to $m\tau $ present a better accuracy. For the values
considered above, the lines (a), (b) and (c) correspond to $eE\tau ^{2}=100$,%
$\ 2500$ and $10000$, respectively.

\subsubsection{Total numbers\label{Sec3.1.2}}

In this section we estimate the total number of pairs created $N^{\mathrm{cr}%
}$ and the vacuum-vacuum transition probability $P_{v}$ (\ref{22.2}) in the
slowly varying approximation (\ref{22.3}). For $t$-electric potential steps,
the total number of pairs created is proportional to the space time volume%
\footnote{%
In Eq. (\ref{3.9.0}), the sum over the quantum numbers $\mathbf{p}$ was
transformed into an integral and the total number of spin polarizations $%
J_{\left( d\right) }=2^{\left[ d/2\right] -1}$ factorizes out from the
density, since $N_{n}^{\mathrm{cr}}$ does not depend on spin variables.}%
\begin{equation}
N^{\mathrm{cr}}=V_{\left( d-1\right) }n^{\mathrm{cr}}\,,\ \ n^{\mathrm{cr}}=%
\frac{J_{\left( d\right) }}{\left( 2\pi \right) ^{d-1}}\int d\mathbf{p}%
N_{n}^{\mathrm{cr}}\,,  \label{3.9.0}
\end{equation}%
so that it is reduced to the calculation of the total density of pairs
created $n^{\mathrm{cr}}$. Similarly to other exactly solvable cases (see
Refs.~\cite{slow-var17,AdoGavGit17}), to evaluate the total density within the
slowly varying configuration (\ref{22.3}), one may restrict to the
calculation of its dominant contribution $\tilde{n}^{\mathrm{cr}}$,
characterized by an integration domain of the quantum numbers $\mathbf{p}$
in which the density is linear in the total increment of the longitudinal
kinetic momentum $\Delta U=e\left\vert A_{x}\left( +\infty \right)
-A_{x}\left( -\infty \right) \right\vert $. We conveniently denote this
domain by $\Omega $ and express the dominant contribution by $\tilde{n}^{%
\mathrm{cr}}$, so that the density is approximately given by%
\begin{equation}
n^{\mathrm{cr}}\approx \tilde{n}^{\mathrm{cr}}=\frac{J_{\left( d\right) }}{%
\left( 2\pi \right) ^{d-1}}\int_{\mathbf{p}\in \Omega }d\mathbf{p}N_{n}^{%
\mathrm{cr}}\,,  \label{3.9.1}
\end{equation}

The above analysis\ shows that dominant contributions for mean numbers of
created particles by a slowly varying field are formed in ranges of large
longitudinal kinetic momenta, restricted values to $\mathbf{p}_{\perp }$,
and have the asymptotic forms (\ref{22.13}). In this case, $\Omega $\ is
realized as%
\[
\Omega :\left\{ \pi _{\perp }-eE\tau _{1}\leq p_{x}\leq -\pi _{\perp
}+eE\tau _{2}\,,\ \ \sqrt{\lambda }<K_{\perp }\right\} \,, 
\]%
so that the dominant density may be expressed as follows: 
\begin{eqnarray}
\tilde{n}^{\mathrm{cr}} &=&\frac{J_{\left( d\right) }}{\left( 2\pi \right)
^{d-1}}\int_{\sqrt{\lambda }<K_{\perp }}d\mathbf{p}_{\perp }\left[ I_{%
\mathbf{p}_{\perp }}^{\left( 1\right) }+I_{\mathbf{p}_{\perp }}^{\left(
2\right) }\right] \,,  \nonumber \\
I_{\mathbf{p}_{\perp }}^{\left( 1\right) } &=&\int_{\pi _{\perp }}^{eE\tau
_{1}}d\pi _{1}e^{-2\pi \nu _{1}^{-}}\,,\ \ I_{\mathbf{p}_{\perp }}^{\left(
2\right) }=\int_{\pi _{\perp }}^{eE\tau _{2}}d\left\vert \pi _{2}\right\vert
e^{-2\pi \nu _{2}^{+}}\,.  \label{22.15}
\end{eqnarray}

Performing two additional change of variables $\lambda s_{1}=2\nu _{1}^{-}$
and $\lambda s_{2}=2\nu _{2}^{+}$ in $I_{\mathbf{p}_{\perp }}^{\left(
1\right) }$ and $I_{\mathbf{p}_{\perp }}^{\left( 2\right) }$, respectively,
and neglecting exponentially small contributions, these integrals can be
rewritten as%
\begin{equation}
I_{\mathbf{p}_{\perp }}^{\left( j\right) }=\int_{1}^{\infty
}ds_{j}F_{j}\left( s_{j}\right) e^{-\pi \lambda s_{j}}\,,\ \ F_{j}\left(
s_{j}\right) =\frac{d\left\vert \pi _{j}\right\vert }{ds_{j}}\,,
\label{22.16}
\end{equation}%
whose superior limits $\lambda s_{j}^{\max }\simeq 4eE\tau _{j}^{2}$ were
extended to infinity for convenience. The leading contributions for
integrals (\ref{22.16}) comes from values near $s_{j}\rightarrow 1$, in
which $F_{j}\left( s_{j}\right) \approx -\left( eE\tau _{j}\right)
/2s_{j}^{3/2}$. Consequently, the leading terms are%
\begin{equation}
I_{\mathbf{p}_{\perp }}^{\left( j\right) }\approx \frac{eE\tau _{j}}{2}%
e^{-\pi \lambda }G\left( \frac{1}{2},\pi \lambda \right) \,,  \label{22.17}
\end{equation}%
where $G\left( \alpha ,x\right) =e^{z}x^{\alpha }\Gamma \left( -\alpha
,x\right) $ and $\Gamma \left( -\alpha ,x\right) $ is the incomplete gamma
function. Neglecting exponentially small contributions, one can extend the
integration limit over $\mathbf{p}_{\bot }$ in Eq.~(\ref{22.15}) from $\sqrt{%
\lambda }<K_{\bot }$ to $\sqrt{\lambda }<\infty $. As a result, the total
density of pairs created (\ref{22.15}) reads%
\begin{equation}
\tilde{n}^{\mathrm{cr}}\approx r^{\mathrm{cr}}\frac{\Delta U_{\mathrm{is}}}{%
eE}\frac{1}{2}G\left( \frac{d-1}{2},\frac{\pi m^{2}}{eE}\right) \,,\ \ r^{%
\mathrm{cr}}=\frac{J_{\left( d\right) }\left( eE\right) ^{\frac{d}{2}}}{%
\left( 2\pi \right) ^{d-1}}\exp \left( -\frac{\pi m^{2}}{eE}\right) \,.
\label{22.18}
\end{equation}%
Here $r^{\mathrm{cr}}$ is rate of pair creation and $\Delta U_{\mathrm{is}%
}=e\left\vert A\left( +\infty \right) -A\left( -\infty \right) \right\vert
=eE\left( \tau _{1}+\tau _{2}\right) $ denotes the total increment of the
longitudinal kinetic momentum for the inverse square electric field. Under
these approximations, the vacuum-vacuum transition probability (\ref{22.2})
has the form%
\begin{eqnarray}
&&P_{v}\approx \exp \left( -\mu N^{\mathrm{cr}}\right) \,,\ \ \mu
=\sum_{l=0}^{\infty }\frac{\left( -1\right) ^{\left( 1-\kappa \right)
l/2}\epsilon _{l+1}}{\left( l+1\right) ^{d/2}}\,\exp \left( -l\pi \frac{m^{2}%
}{eE}\right) ,  \nonumber \\
&&\epsilon _{l}=G\left( \frac{d-1}{2},l\pi \frac{m^{2}}{eE}\right) G\left( 
\frac{d-1}{2},\frac{\pi m^{2}}{eE}\right) ^{-1}\,,  \label{22.19}
\end{eqnarray}%
in leading-order approximation. It should be noted that Eqs. (\ref{22.18})
and (\ref{22.19}) can be equivalently obtained from universal forms for
slowly varying $t$-electric potential steps given by in Ref.~\cite{slow-var17}%
. Explicitly, one can use the universal form of the dominant density given
by Eq. (3.6) in Ref.~\cite{slow-var17},%
\begin{equation}
\tilde{n}^{\mathrm{cr}}\approx \sum_{j}\tilde{n}_{j}^{\mathrm{cr}},\ \ 
\tilde{n}_{j}^{\mathrm{cr}}=\frac{J_{\left( d\right) }}{\left( 2\pi \right)
^{d-1}}\int_{t\in D_{j}}dt\left[ eE_{j}\left( t\right) \right] ^{d/2}\exp %
\left[ -\frac{\pi m^{2}}{eE_{j}\left( t\right) }\right] \,,  \label{22.19.1}
\end{equation}%
to show that Eq. (\ref{22.19.1}) coincides with Eq. (\ref{22.18}) after a
convenient change of variables. Here $D_{j}=\left\{ D_{1}=\mathrm{I},\ D_{2}=%
\mathrm{II}\right\} $ denotes the integration domain for each interval of
definition of the electric field (\ref{2.0}). This is one more independent
confirmation of the universal form for the total number of pairs created
from the vacuum by slowly varying backgrounds.

The representation given by Eq. (\ref{22.18}) is particularly useful to
compare the present results with another exactly solvable examples, for
instance a $T$-constant electric field \cite{BagGitS75,GavGit96} and a peak
electric field \cite{AGG16}, whose dominant densities are proportional
to the corresponding total increment of the longitudinal kinetic momentum in
the slowly varying regime. Recalling the definitions of the $T$-constant
electric field and the peak electric field \cite%
{GavGit96,AGG16,AdoGavGit17}%
\begin{eqnarray}
&\mathrm{(i)\ }E\left( t\right) =&E\,,\ \ t\in \left[ -T/2,T/2\right] \,, 
\nonumber \\
&\mathrm{(ii)\ }E\left( t\right) =&E\left\{ 
\begin{array}{ll}
e^{k_{1}t}\,, & t\in \mathrm{I} \\ 
e^{-k_{2}t}\,, & t\in \mathrm{II}%
\end{array}%
\right. \,,  \label{3.16}
\end{eqnarray}%
as well as their corresponding dominant densities of pair creation in the
slowly varying approximation%
\begin{eqnarray}
&\mathrm{(i)\ }\tilde{n}^{\mathrm{cr}}=&r^{\mathrm{cr}}\frac{\Delta U_{%
\mathrm{T}}}{eE}\,,\ \ \Delta U_{\mathrm{T}}=eET\,,  \nonumber \\
&\mathrm{(ii)\ }\tilde{n}^{\mathrm{cr}}=&r^{\mathrm{cr}}\frac{\Delta U_{%
\mathrm{p}}}{eE}G\left( \frac{d}{2},\frac{\pi m^{2}}{eE}\right) \,,\ \
\Delta U_{\mathrm{p}}=eE\left( k_{1}^{-1}+k_{2}^{-1}\right) \,,  \label{3.17}
\end{eqnarray}%
one can establish relations among these fields by which they are equivalent
in pair production. For example, equating dominant densities for a given
amplitude $E$ and same longitudinal kinetic momentum increments $\Delta U_{%
\mathrm{p}}=\Delta U_{\mathrm{T}}$, we have shown in \cite%
{BagGitS75,slow-var17,AdoGavGit17} that the peak electric field is
equivalent to a $T$-constant electric field in pair production, provided
that it acts on the vacuum over an effective time duration%
\begin{equation}
T_{\mathrm{eff}}=\left( k_{1}^{-1}+k_{2}^{-1}\right) G\left( \frac{d}{2},%
\frac{\pi m^{2}}{eE}\right) \,,  \label{3.17b}
\end{equation}%
(cf. Eq. (3.26) in \cite{AGG16}). By definition, $T_{\mathrm{eff}}=T$
for a $T$-constant field. In other words, a $T$-constant field acting over
the time interval $T=T_{\mathrm{eff}}$ is equivalent to the peak electric
field in pair production. Extending these considerations to the case of the
inverse square electric field (\ref{2.0}), we obtain the following effective
time duration%
\begin{equation}
T_{\mathrm{eff}}=\frac{\tau _{1}+\tau _{2}}{2}G\left( \frac{d-1}{2},\frac{%
\pi m^{2}}{eE}\right) \,,  \label{3.18}
\end{equation}%
i.e., a $T$-constant electric field acting on the vacuum over the same
effective time duration $T=T_{\mathrm{eff}}$ is equivalent to the inverse
square electric field (\ref{2.0}) in pair production.

Comparing the effective time duration for the peak electric field (\ref{3.16}%
) and the inverse square electric field (\ref{2.0}), we see that besides
similarities among their exact solutions (in both cases the solutions of the
Dirac equation are proportional to Kummer functions), they also share common
features regarding particle production. These peculiarities suggest a direct
comparison between the peak and inverse square electric fields, assuming
that both acts over the same time duration $T_{\mathrm{eff}}$ and have the
same amplitude $E$, namely%
\begin{equation}
\frac{\tau _{1}+\tau _{2}}{2}=\left( k_{1}^{-1}+k_{2}^{-1}\right) G\left( 
\frac{d}{2},\frac{\pi m^{2}}{eE}\right) G\left( \frac{d-1}{2},\frac{\pi m^{2}%
}{eE}\right) ^{-1}\,,  \label{3.19}
\end{equation}%
so that we obtain a relation between parameters. Let us consider symmetric
fields $\tau _{1}=\tau _{2}=\tau $, $k_{1}=k_{2}=k$. For weak amplitudes, $%
E\ll m^{2}/e$, one can use the asymptotic approximation of the functions
above with large argument $G\left( \alpha ,z\right) \approx z^{-1}e^{-2z}\,$,%
$\ \ z\rightarrow \infty $, to obtain $\tau /2=k^{-1}$. Thus, one may
conclude that a symmetric peak field (cf. Eq. (2.4) in \cite{AGG16})
requires only half of the pulse duration of a symmetric\ inverse square
field (\ref{2.0}) to be equivalent in pair production. Such a relation does
not depend on electron mass, field strength neither space-time dimensions $d$%
. For strong amplitudes $E\gg m^{2}/e$ though, one can restrict to the
leading-order approximation of $G\left( \alpha ,z\right) $ with small
argument, $G\left( \alpha ,z\right) \approx \alpha ^{-1}\,$,$\ \
z\rightarrow 0$, to show that the latter relation does depend on the
space-time dimensions $\tau /2\approx \left( 1-d^{-1}\right) k^{-1}$. As a
result, we see that $\tau \approx k^{-1}$ for the lowest space-time
dimension $d=2$ and conclude that the relation between $\tau $ and $k$
varies within the interval $k^{-1}\leq \tau \leq 2k^{-1}$, for any amplitude 
$E$ or space-time dimensions $d$, provided that both fields acts over the
same effective time duration $T_{\mathrm{eff}}$.

For completeness, it is worth extending the comparison to the level of the
vacuum-vacuum transition probability $P_{v}$. For the peak electric field,
this probability is given by%
\begin{eqnarray}
&&P_{v}\approx \exp \left( -\mu N^{\mathrm{cr}}\right) \,,\ \ \mu
=\sum_{l=0}^{\infty }\frac{\left( -1\right) ^{\left( 1-\kappa \right) l/2}}{%
\left( l+1\right) ^{d/2}}\epsilon _{l+1}^{\mathrm{p}}e^{-\frac{\pi m^{2}}{eE}%
l}\,,  \nonumber \\
&&\epsilon _{l}^{\mathrm{p}}=G\left( \frac{d}{2},\frac{\pi m^{2}}{eE}%
l\right) G\left( \frac{d}{2},\frac{\pi m^{2}}{eE}\right) ^{-1}\,,
\label{3.20}
\end{eqnarray}%
(cf. Eq. (3.23) in \cite{AGG16}) while for the inverse square electric
field it is given by Eq. (\ref{22.19}). Thus we see that $\epsilon _{l}^{%
\mathrm{p}}\approx \epsilon _{l}^{\mathrm{is}}\approx 1$ for strong
amplitudes $E\gg m^{2}/e$ and $\epsilon _{l}^{\mathrm{p}}\approx \epsilon
_{l}^{\mathrm{is}}\approx l^{-1}$ for weak ones $E\ll m^{2}/e$. Accordingly,
one may say that the discrepancy between the time-dependence of both fields
are not essential for the vacuum-vacuum transition probability, provided
that both electric fields have the same amplitude and are equivalent in
production. We stress that this fact is not true for all types of
time-dependent electric fields. For example, the probability $P_{v}$
corresponding to a Sauter-type electric field $E\left( t\right) =E\cosh
^{-2}\left( t/T_{\mathrm{S}}\right) $ \cite{DunHal98,GavGit96,AdoGavGit17}
differs substantially in comparison to the cases under consideration, even
though all of them are equivalent in pair production in what concerns total
numbers of pairs created from the vacuum.

\section{Asymmetric configuration\label{Sec4}}

In the previous section, the inverse square electric field (\ref{2.0}) was
treated in a somewhat symmetrical manner, once the pulses duration $\tau
_{1} $ and $\tau _{2}$ were considered large, approximately equal and with a
fixed ratio $\tau _{1}/\tau _{2}$. Here we supplement the above study with
an essentially asymmetrical configuration for the electric field,
characterized by a very sharp pulse duration in the first interval $\mathrm{I%
}$ while remaining arbitrary in the second interval $\mathrm{II}$. In this
way, the electric field is mainly defined on the positive half-interval. The
present consideration provides insights on switching on or off effects by
inverse square electric fields, as shall be discussed below.

The present configuration is specified by small values to $\tau _{1}$%
\begin{equation}
0\leq eE\tau _{1}^{2}\ll \min \left( 1,\frac{m^{2}}{eE}\right) \,,
\label{4.00}
\end{equation}%
which includes, as a particular case, the inverse square decreasing electric
field%
\begin{equation}
E\left( t\right) =E\left( 1+t/\tau _{2}\right) ^{-2}\,,\ A_{x}\left(
t\right) =E\tau _{2}\left[ \left( 1+t/\tau _{2}\right) ^{-1}-1\right] \,,
\label{4.0}
\end{equation}%
when $eE\tau _{1}^{2}=0$. Besides the condition (\ref{4.00}), we are
interested in a slowly varying configuration for $t\in \mathrm{II}$, which
means that the pulse duration scales $\tau _{1},\tau _{2}$ obeys additional
conditions%
\begin{equation}
eE\tau _{2}^{2}\gg K_{\perp }^{2}\gg \max \left( 1,\frac{m^{2}}{eE}\right)
\,,\ \ \sqrt{eE}\tau _{1}\sqrt{eE}\tau _{2}\ll 1\,.  \label{4.3}
\end{equation}%
The rightmost inequality implies that the parameter\textrm{\ }$\sqrt{eE}\tau
_{1}$\textrm{\ }is very small, so that the contribution from the first
interval $t\in I$\ is negligible for particle creation. To see that, it is
sufficient to compare the $g$-coefficient $g\left( _{-}|^{+}\right) $ given
by Eq. (\ref{21.15}) in the limit $\sqrt{eE}\tau _{1}\rightarrow 0$ with the
one computed directly for the inverse square decreasing electric field (\ref%
{4.0}). To this end, one may repeat the same considerations as in Sec. \ref%
{Sec2} and take into account that the only essential difference between the
fields (\ref{2.0}) and (\ref{4.0}) lies on the interval \textrm{I}, whose
exact solutions of Eq. (\ref{s2}) are now plane waves,%
\begin{equation}
\ _{\pm }\varphi _{n}\left( t\right) =\ _{\pm }\mathcal{N}e^{\mp i\omega
_{0}t}\,,\ \ \omega _{0}=\sqrt{\mathbf{p}^{2}+m^{2}}\,,\ \ t\in \mathrm{I\,}.
\label{s21.0}
\end{equation}%
Calculating the corresponding normalization constants$\ _{\pm }\mathcal{N}$
for this case one obtains, after some elementary manipulations, the
following form for the $g$-coefficient $g\left( _{-}|^{+}\right) $%
\begin{eqnarray}
&&g\left( _{-}|^{+}\right) =\kappa e^{i\pi \left( 1+\chi \right)
/4}e^{i\theta _{2}}\sqrt{\frac{q_{0}^{+\chi }}{\omega _{2}q_{2}^{-\chi
}\omega _{0}}}\left( 2\omega _{2}\tau _{2}\right) ^{\left( 1+\chi \right)
/2}e^{-\frac{\pi \nu _{2}^{+}}{2}}\Delta _{0}\left( 0\right) \,,  \nonumber
\\
&&\Delta _{0}\left( t\right) =\frac{1}{2}\omega _{0}\Psi \left(
a_{2},c_{2};z_{2}\right) +f_{2}^{-}\left( t\right) \,,\ \ q_{0}^{+\chi
}=\left. q_{1}^{+\chi }\right\vert _{\tau _{1}=0}\,,\ \ \theta _{2}=\left.
\theta _{+}\right\vert _{\tau _{1}=0}\,.  \label{s21.1}
\end{eqnarray}%
It can be readily seen that Eq. (\ref{s21.1}) is a particular case of Eq. (%
\ref{21.15}) when $\tau _{1}=0$. To demonstrate that, one has to select a
particular value to $\chi $ since the Whittaker functions has different
limiting forms as $z_{1}\rightarrow 0$ for each chosen $\mu _{1}$. For
example, let us consider the Fermi case with the choice $\chi =-1$. Thus,
using the approximations $\mu _{1}\approx 1/2$, $\kappa _{1}\approx 0$ and
the limiting form given by Eq. \ref{ap2.5} in Appendix \ref{App1}, we obtain%
\begin{equation}
W_{0,\frac{1}{2}}\left( e^{-i\pi }z_{1}\right) \approx 1\,,\ \ \frac{d}{%
dz_{1}}W_{0,\frac{1}{2}}\left( e^{-i\pi }z_{1}\right) \approx \frac{1}{2}%
\,,\ \ z_{1}\rightarrow 0\,,  \label{3.1}
\end{equation}%
and conclude that Eq. (\ref{21.15}) coincides with the coefficient (\ref%
{s21.1}) under the choices $\kappa =+1$ and $\chi =-1$ in leading-order
approximation\footnote{%
A similar demonstration can be carried out for the Klein-Gordon\ case.}. As
a result, the influence from the first interval $\mathrm{I}$ appears only as
next-to-leading order corrections, which means that we can study pair
creation by the inverse square decreasing electric field (\ref{4.0}) rather
than by the inverse square field (\ref{2.0}) with $eE\tau _{1}^{2}$ obeying
the conditions (\ref{4.00}), in leading-order approximation. Therefore,
without loss of generality, we shall study particle creation by the field (%
\ref{4.0}). Note that from the\textbf{\ }property of the differential mean
numbers $N_{n}^{\mathrm{cr}}$ under the exchanges $p_{x}\rightleftarrows
-p_{x}$ and $\tau _{1}\rightleftarrows \tau _{2}$, the present discussion
can be easily generalized to a configuration in which the field is arbitrary
during the first interval $\mathrm{I}$ but sharp during the second interval $%
\mathrm{II}$.

As discussed previously, only a limited interval of values of the quantum
numbers $\mathbf{p}$ contributes significantly to the differential mean
numbers $N_{n}^{\mathrm{cr}}$. Accordingly, the most significant
contribution comes from finite values to the perpendicular momenta $\mathbf{p%
}_{\perp }$, satisfying $\sqrt{\lambda }<K_{\perp }$ in which $K_{\perp }$
is any number within the interval $eE\tau _{2}^{2}\gg K_{\perp }^{2}\gg \max
\left( 1,m^{2}/eE\right) $. As for the longitudinal momentum $p_{x}$, the
most important contribution comes from the range%
\begin{equation}
\left( \tilde{c}\right) \ -\sqrt{eE}\tau _{2}\left( 1-\tilde{\Upsilon}%
_{2}\right) \leq \frac{\pi _{2}}{\sqrt{eE}}<-\sqrt{\lambda }\,,  \label{3.2}
\end{equation}%
where $0<\tilde{\Upsilon}_{2}\ll 1$ is a number such that $p_{x}/\sqrt{eE}$
is finite, $\min \left( p_{x}/\sqrt{eE}\right) =\tilde{\Upsilon}_{2}\sqrt{eE}%
\tau _{2}$. In this range, the auxiliary variable $\mathcal{Z}_{2}$ defined
in Eq. (\ref{22.6}) is considered large, since $\eta _{2}\approx 1-\tilde{%
\Upsilon}_{2}$. Using the asymptotic approximation of the CHF given by the
first line of Eq. (\ref{ap10}) in Appendix \ref{App1}, we find that the
differential mean number of particles created takes the form%
\begin{equation}
N_{n}^{\mathrm{cr}}\approx \exp \left( -2\pi \nu _{2}^{+}\right) \,.
\label{3.4}
\end{equation}%
This result is valid for Fermions and Bosons.

Besides the range above, there are two additional ones%
\begin{eqnarray}
&&\left( \tilde{a}\right) \ -\frac{\tilde{\delta}_{2}}{\sqrt{2}}-\sqrt{eE}%
\tau _{2}\leq \frac{\pi _{2}}{\sqrt{eE}}\leq \frac{\tilde{\delta}_{2}}{\sqrt{%
2}}-\sqrt{eE}\tau _{2}\,,  \nonumber \\
&&\left( \tilde{b}\right) \ -\sqrt{eE}\tau _{2}+\frac{\tilde{\delta}_{2}}{%
\sqrt{2}}<\frac{\pi _{2}}{\sqrt{eE}}<-\sqrt{eE}\tau _{2}\left( 1-\tilde{%
\Upsilon}_{2}\right) \,,  \label{3.6}
\end{eqnarray}%
in which $0<\tilde{\delta}_{2}\ll 1$ is a small number. In the first
interval $\left( \tilde{a}\right) $, $\eta _{2}\approx 1-\tilde{\delta}_{2}/%
\sqrt{2eE}\tau _{2}$ and $\left\vert \mathcal{Z}_{2}\right\vert \lesssim 
\tilde{\delta}_{2}$ is considered small so that one can use the asymptotic
approximation given by Eq. (\ref{ap6}) in Appendix \ref{App1} to show that
the mean number of electron/positron pairs are given by%
\begin{eqnarray}
&&N_{n}^{\mathrm{cr}}\approx \frac{1}{2}\left[ 1-\sqrt{1-\exp \left( -2\pi
\nu _{2}^{+}\right) }\cos \theta \right] \,,  \nonumber \\
&&\theta =\frac{\pi }{4}+\arg \Gamma \left( \frac{i\nu _{2}^{+}}{2}\right)
-\arg \Gamma \left( \frac{1}{2}+\frac{i\nu _{2}^{+}}{2}\right) \,.
\label{3.7}
\end{eqnarray}%
A similar expression can be obtained for Klein-Gordon particles. In the
interval $\left( \tilde{b}\right) $, the auxiliary variable $\mathcal{Z}_{2}$
is finite. Thus, the uniform asymptotic approximation (\ref{ap1}) can be
used to simplify the CHF $\Psi \left( a_{2},c_{2};z_{2}\right) $.

\begin{figure}[th!]
\begin{center}
\includegraphics[scale=0.48]{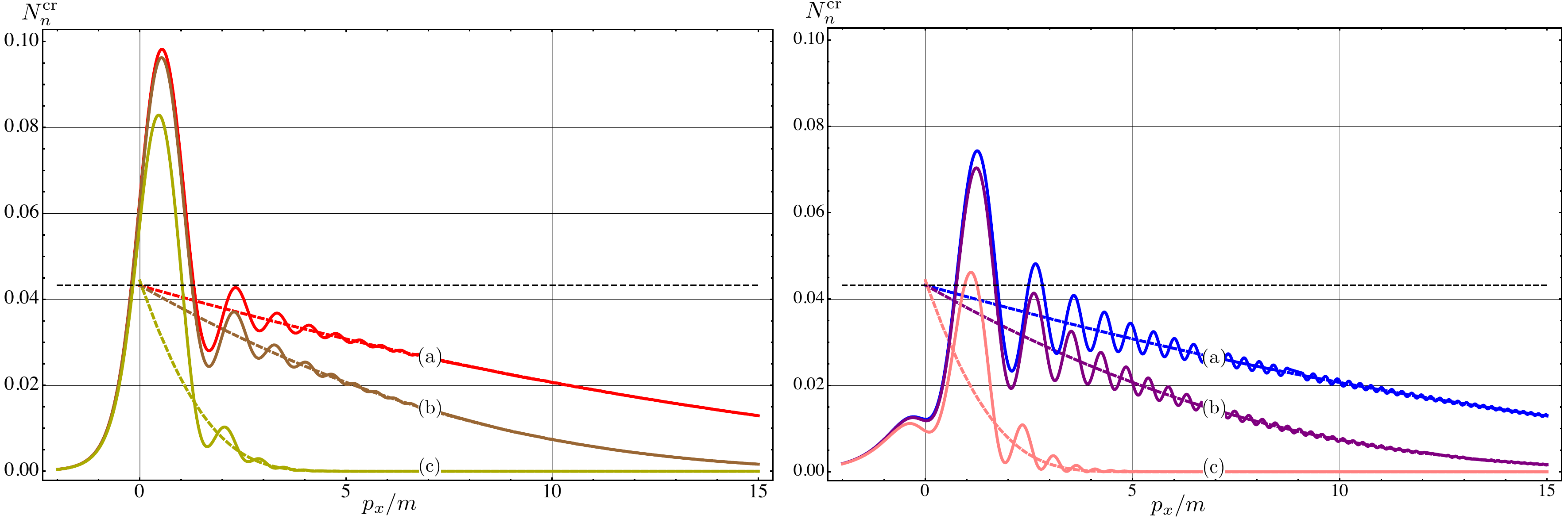}
\end{center}
\caption{(Color online) Differential
mean number $N_{n}^{\mathrm{cr}}$ of Fermions (left panel) and Bosons (right
panel) created from the vacuum by an inverse square decreasing electric
field (\protect\ref{4.0}). The exact differential mean numbers $N_{n}^{\mathrm{cr}}$
given by the absolute squared value of Eq. (\protect\ref{s21.1}) are
represented by solid lines while the asymptotic approximation (\protect\ref%
{3.4}) are represented by dashed lines. The lines labelled with $(\mathrm{a}%
) $,\ $\left( \mathrm{b}\right) $ and $\left( \mathrm{c}\right) $, refers to 
$m\protect\tau _{2}=100$, $50$ and $10$, respectively. In both plots, $E=E_{%
\mathrm{c}}$ and the horizontal dahed line denotes the uniform distribution $%
e^{-\protect\pi \protect\lambda }$ which, in this case, is $e^{-\protect\pi %
} $.}
\label{Fig4}
\end{figure}

\begin{figure}[th!]
\begin{center}
\includegraphics[scale=0.34]{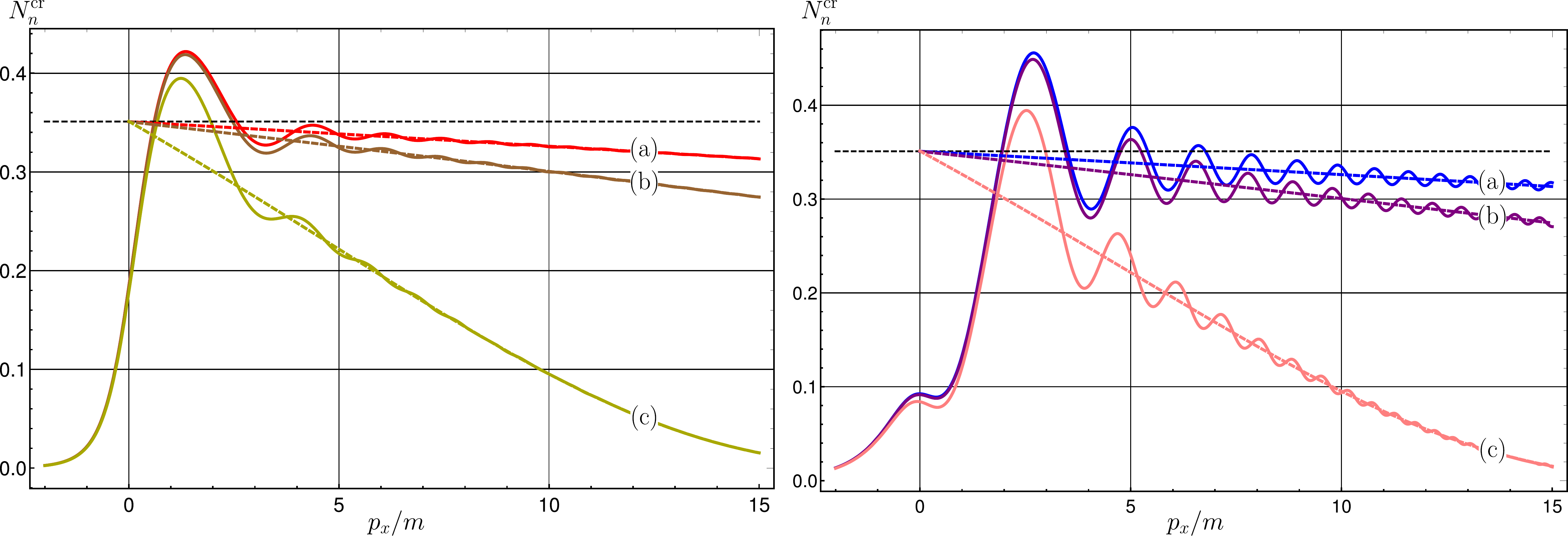}
\end{center}
\caption{(Color online) Differential
mean number $N_{n}^{\mathrm{cr}}$ of Fermions (left panel) and Bosons (right
panel) created from the vacuum by an inverse square decreasing electric
field (\protect\ref{4.0}). The exact differential mean numbers $N_{n}^{\mathrm{cr}}$
given by the absolute squared value of Eq. (\protect\ref{s21.1}) are
represented by solid lines while the asymptotic approximation (\protect\ref%
{3.4}) are represented by dashed lines. The lines labelled with $(\mathrm{a}%
) $,\ $\left( \mathrm{b}\right) $ and $\left( \mathrm{c}\right) $, refers to 
$m\protect\tau _{2}=100$, $50$ and $10$, respectively. In both plots, $E=3E_{%
\mathrm{c}}$ and the horizontal dahed line denotes the uniform distribution $%
e^{-\protect\pi \protect\lambda }$ which, in this case, is $e^{-\protect\pi %
/3}$.}
\label{Fig5}
\end{figure}

The approximation (\ref{3.4}) tends to the uniform distribution $e^{-\pi
\lambda }$ in leading-order approximation for sufficiently large and
negative longitudinal kinetic momentum $\pi _{2}$, satisfying $\pi _{2}/%
\sqrt{eE}\gtrsim -\sqrt{eE}\tau _{2}$. This result clearly differs from the
approximation (\ref{3.7}), obtained from the exact mean number (absolute
squared value of Eq. (\ref{s21.1})) for the same interval of the
longitudinal kinetic momentum $\pi _{2}$. Such a discrepancy is due to the
asymmetrical time-dependence of the electric field, once the asymptotic
forms agree mutually as $p_{x}$ vary over intervals discussed in Sec. \ref%
{Sec3.1.1} for the inverse square electric field (\ref{2.0}), whose temporal
dependence is almost symmetric. This indicates a clear difference in how the
differential mean numbers $N_{n}^{\mathrm{cr}}$ of pairs created by an
inverse square decreasing electric field (\ref{4.0}) are distributed over
the quantum numbers when compared to inverse square electric field (\ref{2.0}%
) in the range of large $\pi _{2}$, although both mean numbers agrees for
finite or sufficiently large $\pi _{2}$, as it follows from the asymptotic
forms (\ref{22.13}) and (\ref{3.4}). To explore these peculiarities, we
present in Figs. \ref{Fig4} and \ref{Fig5} the exact mean number of pairs
created from the vacuum $N_{n}^{\mathrm{cr}}$ given by the absolute squared
value of Eq. (\ref{s21.1}) and the asymptotic approximation (\ref{3.4}), as
a function of the longitudinal momentum $p_{x}$ for the same values of the
of $\tau _{2}$ and $E$ considered in Sec. \ref{Sec3.1}. As before, we set $%
\mathbf{p}_{\perp }=0$ and select the system in which $\hslash =c=m=1$.%

According to the graphs above, the mean number of pairs created $N_{n}^{%
\mathrm{cr}}$ tends to the uniform distribution $e^{-\pi \lambda }$ as $\tau
_{2}$ increases. This is not unexpected since the inverse square decreasing
electric field (\ref{4.0}) tends to a constant field in the limit $\tau
_{2}\rightarrow \infty $; hence the exact mean numbers should approach to
the uniform distribution as $\tau _{2}$ increases. Moreover, for $\tau _{2}$
fixed, the mean numbers approach to the uniform distribution as the
amplitude $E$ increases, as it can be seen comparing the results from Fig. %
\ref{Fig4} with those of Fig. \ref{Fig5}. This is related with the extend of
the dimensionless parameter $eE\tau _{2}^{2}$ and its comparison to the
threshold value $\max \left( 1,m^{2}/eE\right) $: the greater the parameter $%
eE\tau _{2}^{2}$ is in comparison to $\max \left( 1,m^{2}/eE\right) $, the
closer the mean numbers $N_{n}^{\mathrm{cr}}$ approach to the uniform
distribution $e^{-\pi \lambda }$, which is characteristic to constant
electric fields (or a $T$-constant electric field varying slowly in time).

For $p_{x}$ sufficiently large, the exact results agree with the asymptotic
approximation given by Eq. (\ref{3.4}), as it can be observed comparing
solid and dashed lines. This is a consequence of the fact that there are
values of finite longitudinal kinetic momentum $\pi _{2}$ ($p_{x}$ finite,
range $\left( \tilde{c}\right) $) in which the mean numbers tend to the
asymptotic forms (\ref{3.4}) in slowly varying regime. On the other hand, in
the range of sufficiently small $p_{x}$ (or sufficiently large $\pi _{2}$),
there are deviations between the exact mean numbers and the asymptotic
approximations. Such deviations are expected and usually occurs in the range
of small $p_{x}$, as in the case inverse square electric field (\ref{2.0}),
displayed in Figs. \ref{Fig1} and \ref{Fig2}, or peak electric field \cite%
{AFGG17}, displayed in Fig. 4{\Large \ }of this reference. We conclude that
the approximation of slowly varying regime does not apply uniformly
throughout all values of $p_{x}$ for values of $eE\tau _{2}^{2}$ considered
in the plots above. To be applicable uniformly, larger values of parameters
are needed.

The most striking feature of the results displayed above is the presence of
oscillations, an absent feature in the case of the inverse square electric
field (\ref{2.0}); compare Figs. \ref{Fig1}, \ref{Fig2} with \ref{Fig4}.
These oscillations are consequences of an \textquotedblleft
abrupt\textquotedblright\ switching on process near $t=0$ and frequently
occurs in these cases, as reported recently by us in \cite{AdoFerGavGit18}.
In this work, oscillations around the uniform distribution were found and
discussed for the case of a $T$-constant electric field (that switches-on
and off \textquotedblleft abruptly\textquotedblright\ at definite time
instants) and an electric field composed by independent intervals, one
exponentially increasing, another constant over the duration $T$ and a third
one exponentially decreasing. This is an universal feature of
\textquotedblleft abrupt\textquotedblright\ switching on or off processes.
Moreover, comparing the results displayed in Figs. \ref{Fig4} and \ref{Fig5}
we conclude that the oscillations decrease in magnitude as the parameter $%
eE\tau _{2}^{2}$ increases. As a result, the mean numbers are expected to
become \textquotedblleft rectangular\textquotedblright\ in the limit $eE\tau
_{2}^{2}\rightarrow \infty $.

From the above considerations and the approximations given by Eq. (\ref{3.4}%
), we conclude that the dominant density of pairs created $\tilde{n}^{%
\mathrm{cr}}$ (\ref{3.9.1}) can be expressed as%
\begin{equation}
\tilde{n}^{\mathrm{cr}}\approx r^{\mathrm{cr}}\frac{\tau _{2}}{2}G\left( 
\frac{d-1}{2},\frac{\pi m^{2}}{eE}\right) \,.  \label{3.9}
\end{equation}%
We see that $\tilde{n}^{\mathrm{cr}}$ given by Eq.~(\ref{3.9}) can be
obtained from Eq. (\ref{22.18}) setting $\tau _{1}\rightarrow 0$. The
vacuum-vacuum transition probability has the form $P_{v}=\exp \left( -\mu N^{%
\mathrm{cr}}\right) $, with $\mu $ given by Eq. (\ref{22.19}).

\section{Switching on and off by inverse square electric fields\label{Sec5}}

As an application of the above results, we consider in this section an
electric field of special configuration in which inverse square increasing
and decreasing electric fields simulate switching on and off processes. This
consideration allow us to compare effects with recent results \cite%
{AdoFerGavGit18}, in which a composite electric field of similar form was
regarded to study the influence of switching on and off processes in the
vacuum.

The field under consideration is composed by three independent intervals,
switching on over the first interval $t\in \mathrm{I}=\left( -\infty
,t_{1}\right) $, remains constant over the intermediate interval $t\in 
\mathrm{Int}=\left[ t_{1},t_{2}\right] $ and switching off over the last
interval $t\in \mathrm{II}=\left( t_{2},+\infty \right) $. The field has the
form%
\begin{equation}
E\left( t\right) =E\left\{ 
\begin{array}{ll}
\left[ 1-\left( t-t_{1}\right) /\tau _{1}\right] ^{-2}\,, & t\in \mathrm{I}%
\,, \\ 
1\,, & t\in \mathrm{Int\,}, \\ 
\left[ 1+\left( t-t_{2}\right) /\tau _{2}\right] ^{-2}\,, & t\in \mathrm{II\,%
},%
\end{array}%
\right. \,,  \label{c.1}
\end{equation}%
and, correspondingly, its potential is%
\begin{equation}
A_{x}\left( t\right) =E\left\{ 
\begin{array}{ll}
\tau _{1}-t_{1}-\tau _{1}\left[ 1-\left( t-t_{1}\right) /\tau _{1}\right]
^{-1}\,, & t\in \mathrm{I}\,, \\ 
-t\,, & t\in \mathrm{Int\,}, \\ 
-\tau _{2}-t_{2}+\tau _{2}\left[ 1+\left( t-t_{2}\right) /\tau _{2}\right]
^{-1}\,, & t\in \mathrm{II\,},%
\end{array}%
\right.  \label{c.2}
\end{equation}%
where $t_{1}<0$ and $t_{2}>0$ are fixed time instants.

The existence of an intermediate interval in which the field is constant, $%
t\in \mathrm{Int}$, does not change the classification of particle and
antiparticle states at asymptotic times given by Eq. (\ref{21.10}). However,
it introduces certain modifications on the variables and parameters of the
Whittaker functions, namely\footnote{%
Exclusively in this section, the variables $z_{j}\left( t\right) $ and
parameters $\kappa _{j}$ are defined according to Eqs. (\ref{c.3}), (\ref%
{c.4}) and should not be confused with the previous definitions, given by
Eqs. (\ref{21.1}) and (\ref{21.4}).}%
\begin{eqnarray}
&&z_{1}\left( t\right) =2i\omega _{1}\tau _{1}\left[ 1-\left( t-t_{1}\right)
/\tau _{1}\right] \,,\ \ t\in \mathrm{I}\,,  \label{c.3} \\
&&z_{2}\left( t\right) =2i\omega _{2}\tau _{2}\left[ 1+\left( t-t_{2}\right)
/\tau _{2}\right] \,,\ \ t\in \mathrm{II}\,,  \nonumber
\end{eqnarray}%
and%
\begin{equation}
\kappa _{j}=-\left( -1\right) ^{j}eE\tau _{j}^{2}\frac{\Pi _{j}}{\omega _{j}}%
\,,\ \ \Pi _{j}=p_{x}-eE\left[ t_{j}+\left( -1\right) ^{j}\tau _{j}\right]
\,,\ \ \omega _{j}^{2}=\Pi _{j}^{2}+\pi _{\perp }^{2}\,,  \label{c.4}
\end{equation}%
while the parameters $\mu _{j}$ remains the same as in Eq. (\ref{21.4}).
Hence, exact solutions of Eq. (\ref{s2}) for the intervals $\mathrm{I}$ and $%
\mathrm{II}$ are Whittaker functions, classified according to Eq. (\ref%
{21.10}) with $z_{j}\left( t\right) $ and $\kappa _{j}$ defined by Eqs. (\ref%
{c.3}) and (\ref{c.4}). As for the intermediate interval $t\in \mathrm{Int}$%
, Dirac spinors are proportional to Weber Parabolic Cylinder functions
(WPCFs) \cite{Erdelyi} once the exact solutions of Eq. (\ref{s2}) are
combinations of these functions%
\begin{eqnarray}
&&\varphi _{n}\left( t\right) =b_{+}u_{+}\left( \mathfrak{Z}\right)
+b_{-}u_{-}\left( \mathfrak{Z}\right) \,,\ \ t\in \mathrm{Int\,,}  \nonumber
\\
&&u_{+}\left( \mathfrak{Z}\right) =D_{\nu +\left( \chi -1\right) /2}\left( 
\mathfrak{Z}\right) \,,\ \ u_{-}\left( \mathfrak{Z}\right) =D_{-\nu -\left(
\chi +1\right) /2}\left( i\mathfrak{Z}\right) \,,  \label{c.5}
\end{eqnarray}%
where $b_{\pm }$ are constants and%
\begin{equation}
\mathfrak{Z}\left( t\right) =\left( 1-i\right) \xi \left( t\right) \,,\ \
\xi \left( t\right) =\sqrt{eE}t-\frac{p_{x}}{\sqrt{eE}}\,,\ \ \nu =\frac{%
i\lambda }{2}\,.  \label{c.6}
\end{equation}%
As a result, one may repeat the same steps as described in Sec. \ref{Sec2}
to find the following form to the $g$-coefficients:%
\begin{eqnarray}
g\left( _{-}|^{+}\right) &=&\kappa \sqrt{\frac{q_{1}^{+}}{8eE\omega
_{1}q_{2}^{-}\omega _{2}}}\exp \left[ -\frac{i\pi }{2}\left( \kappa
_{1}+\kappa _{2}-\nu -\frac{\chi }{2}\right) \right]  \nonumber \\
&\times &\left[ \mathcal{F}_{2}^{-}\left( t_{2}\right) \mathcal{G}%
_{1}^{+}\left( t_{1}\right) -\mathcal{F}_{2}^{+}\left( t_{2}\right) \mathcal{%
G}_{1}^{-}\left( t_{1}\right) \right] \,,  \nonumber \\
g\left( ^{+}|_{-}\right) &=&\sqrt{\frac{q_{2}^{-}}{8eE\omega
_{1}q_{1}^{+}\omega _{2}}}\exp \left[ -\frac{i\pi }{2}\left( \kappa
_{1}+\kappa _{2}-\nu -\frac{\chi }{2}\right) \right]  \nonumber \\
&\times &\left[ \mathcal{F}_{1}^{+}\left( t_{1}\right) \mathcal{G}%
_{2}^{-}\left( t_{2}\right) -\mathcal{F}_{1}^{-}\left( t_{1}\right) \mathcal{%
G}_{2}^{+}\left( t_{2}\right) \right] \,.  \label{c.7}
\end{eqnarray}%
Here, $q_{j}^{\pm }=\omega _{j}\pm \chi \Pi _{j}$ and $\mathcal{F}_{j}^{\pm
}\left( t\right) $, $\mathcal{G}_{j}^{\pm }\left( t\right) $\ are
combinations between WPCFs and Whittaker functions%
\begin{eqnarray}
&&\mathcal{F}_{j}^{\pm }\left( t\right) =u_{\pm }\left( \mathfrak{Z}\right) 
\frac{d}{dt}W_{\kappa _{j},\mu _{j}}\left( z_{j}\right) -W_{\kappa _{j},\mu
_{j}}\left( z_{j}\right) \frac{d}{dt}u_{\pm }\left( \mathfrak{Z}\right) \,, 
\nonumber \\
&&\mathcal{G}_{j}^{\pm }\left( t\right) =u_{\pm }\left( \mathfrak{Z}\right) 
\frac{d}{dt}W_{-\kappa _{j},\mu _{j}}\left( e^{-i\pi }z_{j}\right)
-W_{-\kappa _{j},\mu _{j}}\left( e^{-i\pi }z_{j}\right) \frac{d}{dt}u_{\pm
}\left( \mathfrak{Z}\right) \,.  \label{c.8}
\end{eqnarray}

On the basis of the results discussed in Sec. \ref{Sec3.1.1} and previous
studies on the $T$-constant field in the slowly varying regime \cite%
{GavGit96,AdoGavGit17}, we see if the parameters satisfy%
\begin{equation}
\min \left( \sqrt{eE}T,eE\tau _{1}^{2},eE\tau _{2}^{2}\right) \gg \max
\left( 1,\frac{m^{2}}{eE}\right) \,,  \label{c.9}
\end{equation}%
the differential mean number of pairs created acquires the asymptotic form%
\begin{equation}
N_{n}^{\mathrm{cr}}\approx \left\{ 
\begin{array}{ll}
\exp \left( -2\pi \nu _{1}^{-}\right) \,, & \mathrm{for}\ \ p_{x}/\sqrt{eE}%
\leq -\sqrt{eE}T/2\,, \\ 
\exp \left( -\pi \lambda \right) \,, & \mathrm{for}\ \ \left\vert
p_{x}\right\vert /\sqrt{eE}<\sqrt{eE}T/2\,, \\ 
\exp \left( -2\pi \nu _{2}^{+}\right) \,, & \mathrm{for}\ \ p_{x}/\sqrt{eE}%
\geq \sqrt{eE}T/2\,,%
\end{array}%
\right.  \label{c.10}
\end{equation}%
Thus, the total dominant density of pairs created in the slowly varying
regime is a sum of densities%
\begin{equation}
\tilde{n}^{\mathrm{cr}}\approx \tilde{n}_{\mathrm{I}}^{\mathrm{cr}}+\tilde{n}%
_{\mathrm{Int}}^{\mathrm{cr}}+\tilde{n}_{\mathrm{II}}^{\mathrm{cr}}=\left[ T+%
\frac{\tau _{1}+\tau _{2}}{2}G\left( \frac{d-1}{2},\frac{\pi m^{2}}{eE}%
\right) \right] r^{\mathrm{cr}}\,,  \label{c.11}
\end{equation}%
in agreement to the universal form given by Eq. (\ref{22.19.1}) \cite%
{slow-var17}. In cases beyond slowly varying configurations, i.e. when the
conditions (\ref{c.9}) are not satisfied for all parameters, the mean
numbers $N_{n}^{\mathrm{cr}}$ must be studied through the exact expressions
for the $g$-coefficients (\ref{c.7}) according to the definition (\ref{22.1}%
). Hence, in what follows we present mean numbers $N_{n}^{\mathrm{cr}}$ of
pairs created from the vacuum by the composite field (\ref{c.1}) as a
function of the longitudinal momentum $p_{x}$ for some values of the
parameters $\sqrt{eE}\tau _{j}$ and $\sqrt{eE}T$. Moreover, in order to
compare switching on and off effects with an another composite electric
field \cite{AdoFerGavGit18}%
\begin{equation}
E\left( t\right) =E\left\{ 
\begin{array}{ll}
e^{k_{1}\left( t-t_{1}\right) }\,, & t\in \mathrm{I}\,, \\ 
1\,, & t\in \mathrm{Int\,}, \\ 
e^{-k_{2}\left( t-t_{2}\right) }\,, & t\in \mathrm{II\,},%
\end{array}%
\right.  \label{c.12}
\end{equation}%
wherein exponentially increasing and decreasing intervals simulate switching
on and off processes and a $T$-constant field \cite{GavGit96,AdoGavGit17}\
(in which switching on and off processes are absent) we include, in each
graph below, mean numbers of pairs created by the field (\ref{c.12}) and the 
$T$-constant field for some values of the parameters $\sqrt{eE}k_{j}^{-1}$
and $\sqrt{eE}T$.\textbf{\ }As in the previous sections, we set $\mathbf{p}%
_{\perp }=0$ and select the system in which $\hslash =c=m=1$.

\begin{figure}[th!]
\begin{center}
\includegraphics[scale=0.34]{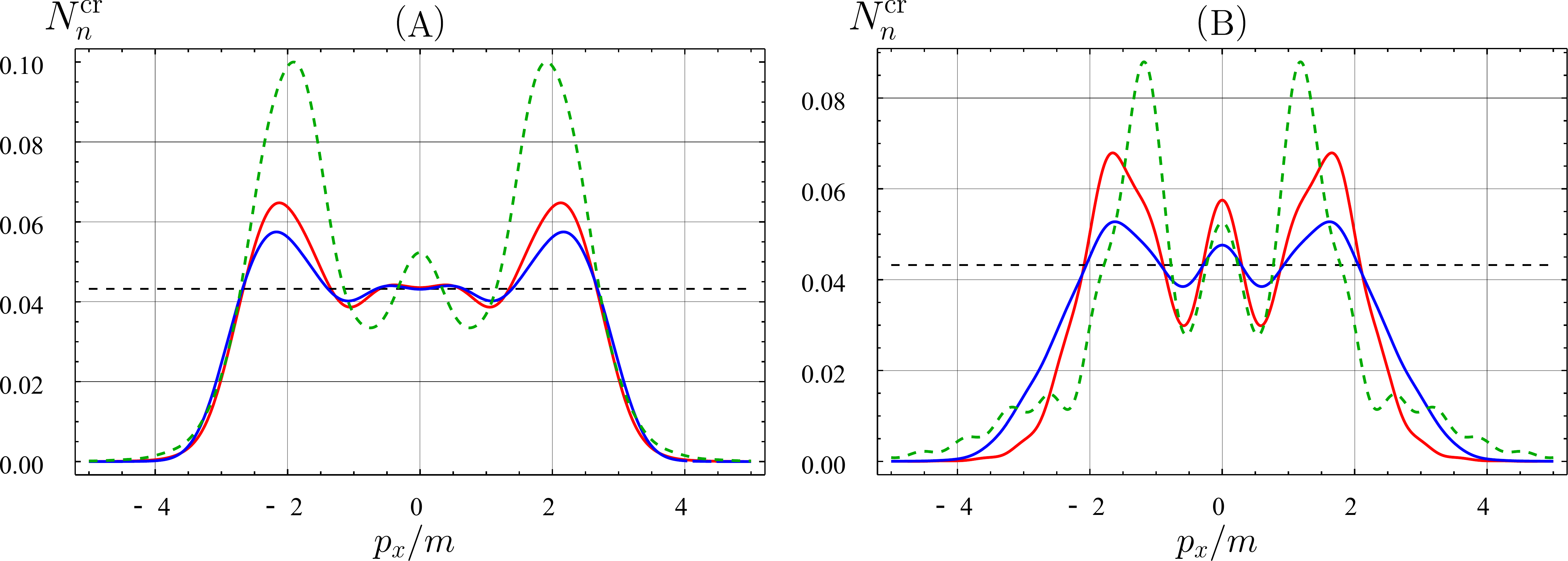}
\end{center}
\caption{(Color online) Differential mean number of
Fermions (A) and Bosons (B) created from the vacuum by electric fields. The
solid red and blue lines refers to the composite fields (\protect\ref{c.1})
and (\protect\ref{c.12}), respectively, with $\protect\tau _{1}=\protect\tau %
_{2}=\protect\tau $ and $k_{1}=k_{2}=k$. The dashed green lines refers to
the $T$-constant field while the horizontal ones denotes the uniform
distribution $e^{-\protect\pi \protect\lambda }$. In both graphs, $m\protect%
\tau =1$, $mk^{-1}=1$, $mT=5$ and $E=E_{\mathrm{c}}$.}
\label{Fig6.1}
\end{figure}

\begin{figure}[th!]
\begin{center}
\includegraphics[scale=0.34]{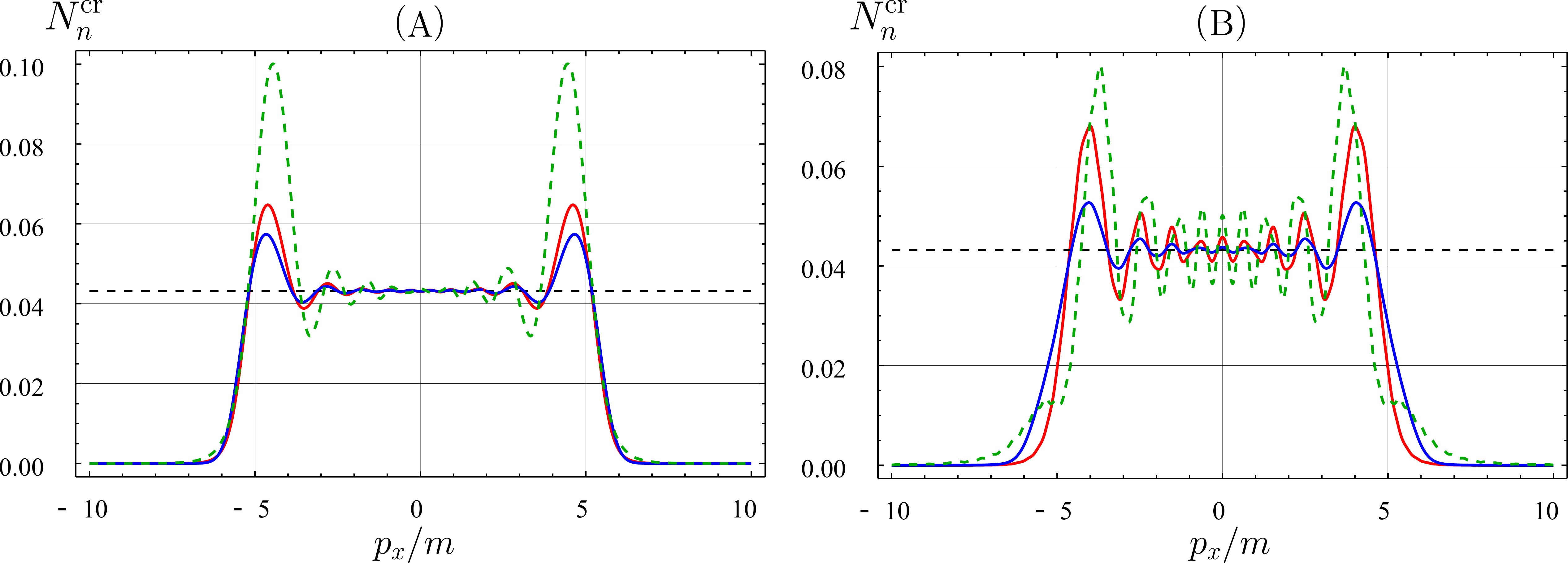}
\end{center}
\caption{(Color online) Differential mean number of Fermions (A) and Bosons
(B) created from the vacuum by electric fields. The solid red and blue lines
refers to the composite fields (\protect\ref{c.1}) and (\protect\ref{c.12}),
respectively, with $\protect\tau _{1}=\protect\tau _{2}=\protect\tau $ and $%
k_{1}=k_{2}=k$. The dashed green lines refers to the $T$-constant field
while the horizontal ones denotes the uniform distribution $e^{-\protect\pi 
\protect\lambda }$. In both graphs, $m\protect\tau =1$, $mk^{-1}=1$, $mT=10$
and $E=E_{\mathrm{c}}$.}
\label{Fig6.2}
\end{figure}

\begin{figure}[th!]
\begin{center}
\includegraphics[scale=0.34]{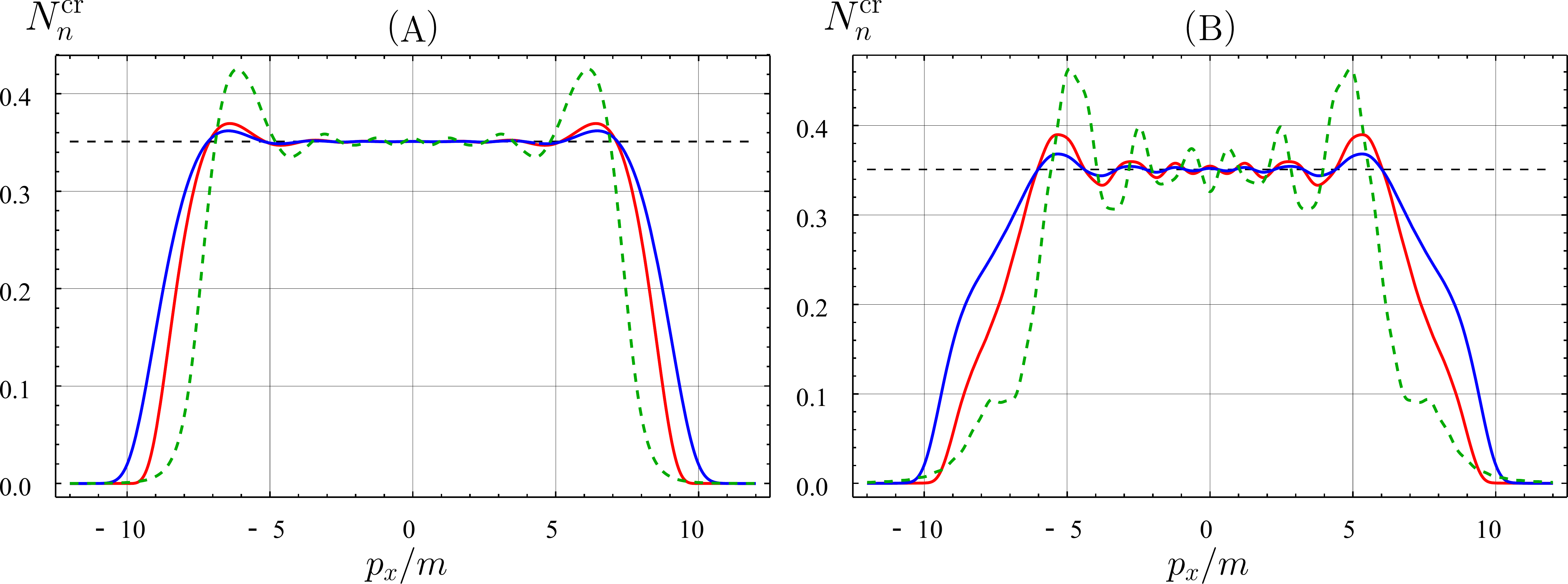}
\end{center}
\caption{(Color online)
Differential mean number of Fermions (A) and Bosons (B) created from the
vacuum by electric fields. The solid red and blue lines refers to the
composite fields (\protect\ref{c.1}) and (\protect\ref{c.12}), respectively,
with $\protect\tau _{1}=\protect\tau _{2}=\protect\tau $ and $k_{1}=k_{2}=k$%
. The dashed green lines refers to the $T$-constant field while the
horizontal ones denotes the uniform distribution $e^{-\protect\pi \protect%
\lambda }$. In both graphs, $m\protect\tau =1$, $mk^{-1}=1$, $mT=5$ and $%
E=3E_{\mathrm{c}}$.}
\label{Fig6.3}
\end{figure}

\begin{figure}[th!]
\begin{center}
\includegraphics[scale=0.34]{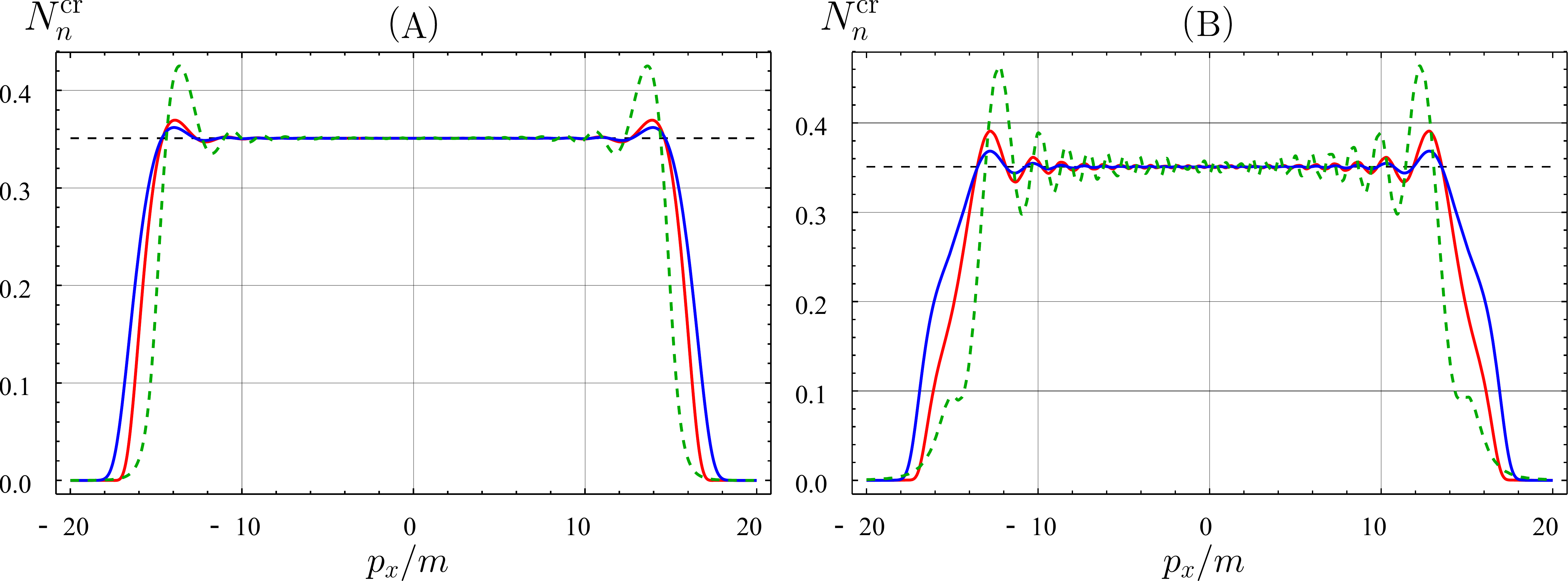}
\end{center}
\caption{(Color online)
Differential mean number of Fermions (A) and Bosons (B) created from the
vacuum by electric fields. The solid red and blue lines refers to the
composite fields (\protect\ref{c.1}) and (\protect\ref{c.12}), respectively,
with $\protect\tau _{1}=\protect\tau _{2}=\protect\tau $ and $k_{1}=k_{2}=k$%
. The dashed green lines refers to the $T$-constant field while the
horizontal ones denotes the uniform distribution $e^{-\protect\pi \protect%
\lambda }$. In both graphs, $m\protect\tau =1$, $mk^{-1}=1$, $mT=10$ and $%
E=3E_{\mathrm{c}}$.}
\label{Fig6.4}
\end{figure}

According to the graphs above, the differential mean numbers oscillate
around the uniform distribution $e^{-\pi \lambda }$, irrespective the
electric field in consideration. This is consistent to asymptotic
predictions for the $T$-constant field, in the sense that the differential
mean numbers $N_{n}^{\mathrm{cr}}$ stabilizes to the uniform distribution $%
e^{-\pi \lambda }$ as soon as $\sqrt{eE}T$ is sufficiently larger than the
characteristic values $\max \left( 1,m^{2}/eE\right) $. Thus, the larger the
value of $\sqrt{eE}T$, the smaller the magnitude of the oscillations. This
explains why oscillations are larger in Fig. \ref{Fig6.1} in which $\sqrt{eE}%
T=5$ in comparison to the ones in Fig. \ref{Fig6.4}, in which $\sqrt{eE}T=10%
\sqrt{3}$. Moreover, one can see that the magnitude of oscillations decrease
if a constant field is accompanied by switching on and off processes;
compare solid and dashed lines. This decrease in the amplitude of the
oscillations is a consequence of smoother switching on and off processes.%
{\Large \ }In the case of the composite field (\ref{c.1}), the mean numbers
are approximated given by the first and third lines of Eq. (\ref{c.10})
while the composite field (\ref{c.12}), $N_{n}^{\mathrm{cr}}\approx \exp
\left( -2\pi \Xi _{1}^{-}\right) $ for $p_{x}/\sqrt{eE}\leq -\sqrt{eE}T/2$
and $N_{n}^{\mathrm{cr}}\approx \exp \left( -2\pi \Xi _{2}^{+}\right) $ for $%
p_{x}/\sqrt{eE}\geq \sqrt{eE}T/2$, in which $\Xi _{j}^{\pm
}=k_{j}^{-1}\left( \sqrt{\tilde{\Pi}_{j}^{2}+\pi _{\perp }^{2}}\pm \tilde{\Pi%
}_{j}\right) $, $\tilde{\Pi}_{j}=p_{x}-\left( -1\right)
^{j}eEk_{j}^{-1}\left( 1+k_{j}T/2\right) $. Accordingly, the exact mean
numbers oscillate around these approximations, whose amplitudes decrease as $%
eE\tau _{j}^{2}$, $eEk_{j}^{-2}$ increases. At last, but not least, we see
that the mean numbers of pairs created by the composite field (\ref{c.12})
oscillate around the uniform distribution less than by the composite field (%
\ref{c.1}), given the same longitudinal kinetic momentum increment of both
switching on and off processes, for all values of the parameters under
consideration. Based on the values chosen for the parameters, we conclude
that the slowly varying regime provides a better approximation to the
composite field (\ref{c.12}) than for the field (\ref{c.1}). However,
assuming the same value for $E$ for both composite fields, it is clear that
for $\tau $ sufficiently larger than $k^{-1}$ (that is, longitudinal kinetic
momentum increment of the inverse square fields is larger than one of
exponential fields), the opposite situation occurs. The composite electric
field (\ref{c.1}) and its peculiarities supply our previous studies \cite%
{AdoFerGavGit18} on the role of switching on and off processes in the vacuum
instability.

\section{Some concluding remarks\label{Sec6}}

In addition to few known exactly solvable cases in QED with external
backgrounds, an inverse square electric field represents one more example
where nonperturbative calculations of particle creation effect can be
performed exactly. We have presented in detail consistent calculations of
zero order quantum effects in the inverse square electric field as well as
in a composite electric field of a special configuration, in which the
inverse square electric field simulates switching on and off processes. In
all these cases we find corresponding \textrm{in} and \textrm{out} exact
solutions of the Dirac and Klein-Gordon equations. Using these solutions, we
calculate differential mean numbers $N_{n}^{\mathrm{cr}}$ of Fermions and
Bosons created from the vacuum. Differential quantities are considered both
exactly and approximately (within the slowly varying regime). In the second
case, we studied these distributions as functions on the particle momenta,
establishing ranges of dominant contributions and finding corresponding
asymptotic representations. In order to be able to compare visually
approximate results with exact ones, we compute and analyze plots of
differential mean numbers $N_{n}^{\mathrm{cr}}$ as functions of $p_{x}$ for
some values to the pulse durations $\tau _{j}$\ and for electric field
magnitude $E$\ equal to the Schwinger's critical value. The asymptotic
representations agree substantially with exact results as the pulse
durations increase. Using the asymptotic representations for differential
quantities, we compute the total number $N^{\mathrm{cr}}$ of created pairs
the probability $P_{v}$ for the vacuum remain the vacuum. The results are
consistent with universal estimates in the locally constant field
approximation. Moreover, comparing the results with dominant densities of
pairs created by the $T$-constant and peak electric fields, we derive an
effective time duration of the inverse square electric field and establish
relations by which they are equivalent in pair production effect. Assuming
that the peak and the inverse square electric fields act on the vacuum over
the same effective time, we relate both fields and conclude that the
relation between their pulses varies as $k^{-1}\leq \tau \leq 2k^{-1}$, for
any amplitude $E$ or space-time dimensions $d$.

To complete the pictures, we consider in Sec. \ref{Sec4} the case of an
asymmetrical configuration, in which the field presents a sharp pulse for $%
t<0$. In the limit $\tau _{1}\rightarrow 0$ the corresponding $g$%
-coefficients are consistent with $g$-coefficients calculated in the
symmetric case. Analyzing plots of exact calculations, we see that the mean
numbers oscillate around their asymptotic approximate values in contrast to
the symmetric case were such oscillations are absent; compare Figs. \ref%
{Fig1}, \ref{Fig2} with Fig. \ref{Fig4}. These oscillations are attributed
to the asymmetrical time dependence of the electric field or, in other
words, to the existence of an \textquotedblleft abrupt\textquotedblright\
switching on process near $t=0$. Moreover, this feature does not depend on
the form of external electric field, they can be observed in other cases,
for instance in $T$-constant electric field (see Figs. \ref{Fig6.1}-\ref%
{Fig6.4}). Thus, we may conclude that the oscillations are universal
features of \textquotedblleft abrupt\textquotedblright\ switching on or off
processes.

Considering an electric field composed by three parts, two of which are
represented by inverse square fields, we calculate relevant $g$-coefficients
for particle creation and discuss approximate expressions for differential
quantities. To understand better switching on and off effects, we compare
the above case with the case where switching on and off configurations have
exponential behavior. Doing this we consider a configuration in which the
duration $T$ of the intermediate $T$ -constant electric fields is greater
than the duration of the characteristic pulses $\tau _{j}$ and $k_{j}^{-1}$.
This configuration allows us to analyze how the differential distributions
differ from their asymptotic form $e^{-\pi \lambda }$. According to Figs. %
\ref{Fig6.1} - \ref{Fig6.4}, we conclude that the way of switching on and
off is essential for application of slowly varying regime approximation. For
example, comparing results in the $T$-constant electric field (dashed lines)
for Fermions with ones for composite fields (solid lines) in Fig. \ref%
{Fig6.3}, we see they are close to results obtained in the slow variation
approximation if parameters of composite fields satisfy the condition $\sqrt{%
eE}T\geq 5\sqrt{3}$, $\sqrt{eE}\tau =\sqrt{eE}k^{-1}=1.$ At the same time,
in the case of a $T$-constant field with $\sqrt{eE}T=5\sqrt{3}$ it is not
true and the corresponding mean numbers $N_{n}^{\mathrm{cr}}$ deviate
substantially from the uniform distribution $e^{-\pi \lambda }$. For Bosons,
one can see that composite fields with $\sqrt{eE}T=5\sqrt{3}$, $\sqrt{eE}%
\tau =\sqrt{eE}k^{-1}=1$ does not allow application of the slow variation
approximation, whereas the condition $\sqrt{eE}T\geq 5\sqrt{3}$ is close to
the threshold condition for composite fields for Fermions. One can also see
that differential quantities are quite sensitive to the form of switching on
and off. For all configurations displayed in Figs. \ref{Fig6.1} - \ref%
{Fig6.4}, we see that exponential switching on and off causes smaller
oscillations around the uniform distribution in comparison to the inverse
square switching on and off.

\section{Acknowledgements}

The authors acknowledge support from Tomsk State University Competitiveness
Improvement Program and the partial support from the Russian Foundation for
Basic Research (RFBR), under the project No. 18-02-00149. T.C.A. also thanks the Advanced Talents Development Program of the Hebei University, project No. 801260201271, for the partial support. D.M.G. is also supported by the Grant No. 2016/03319-6, Funda\c{c}\~{a}o de Amparo \`{a}
Pesquisa do Estado de S\~{a}o Paulo (FAPESP), and permanently by Conselho
Nacional de Desenvolvimento Cient\'{\i}fico e Tecnol\'{o}gico (CNPq), Brazil.

\appendix

\section{Asymptotic representations of special functions\label{App1}}

For $a$ fixed and both $c$ and $z$ large, the CHF $\Psi \left( a,c;z\right) $
admits the following asymptotic approximation \cite{DLMF},%
\begin{eqnarray}
\Psi \left( a,c;z\right) &\approx &c^{-\frac{a}{2}}e^{\frac{\mathcal{Z}^{2}}{%
4}}\mathcal{F}\left( a,c;\eta \right) \,,\ \mathcal{Z}=\left( \eta -1\right) 
\mathcal{W}\sqrt{c}\,,  \nonumber \\
\mathcal{F}\left( a,c;\eta \right) &=&\eta \mathcal{W}^{1-a}D_{-a}\left( 
\mathcal{Z}\right) -\mathcal{R}D_{1-a}\left( \mathcal{Z}\right) \,,
\label{ap1}
\end{eqnarray}%
uniformly valid with respect to the ratio $\eta =z/c\in \left( 0,+\infty
\right) $. Here $\mathcal{W}$, $\mathcal{R}$ are given by%
\begin{equation}
\mathcal{W}=\sqrt{\frac{2\left( \eta -1-\ln \eta \right) }{\left( \eta
-1\right) ^{2}}}\,,\ \ \mathcal{R}=\frac{\eta \mathcal{W}^{1-a}-\mathcal{W}%
^{a}}{\mathcal{Z}}\,,  \label{ap2}
\end{equation}%
and $D_{-a}\left( \mathcal{Z}\right) $, $D_{1-a}\left( \mathcal{Z}\right) $
are Weber's Parabolic Cylinder functions (WPCF) \cite{Erdelyi}. The uniform
asymptotic representation for the derivative has the form%
\begin{equation}
\frac{d\Psi \left( a,c;z\right) }{dz}\approx c^{-\frac{a}{2}}e^{\frac{%
\mathcal{Z}^{2}}{4}}\left( \frac{\eta -1}{2\eta }+\frac{1}{c}\frac{d}{d\eta }%
\right) \mathcal{F}\left( a,c;\eta \right) \,.  \label{ap3}
\end{equation}

When $\left\vert \eta -1\right\vert \rightarrow 0$, $\mathcal{Z}$ is small
so that one can expand the WPCF around $\mathcal{Z}=0$ and subsequently $%
\mathcal{Z}$, $\mathcal{W}$ and $\mathcal{R}$ around $\eta =1$, to show that 
$\Psi \left( a,c;z\right) $ acquires the asymptotic form%
\begin{equation}
\Psi \left( a,c;z\right) \approx c^{-\frac{a}{2}}D_{-a}\left( 0\right) \,,\
\ \left\vert \eta -1\right\vert \rightarrow 0\,.  \label{ap6}
\end{equation}%
For $\left\vert \eta -1\right\vert \rightarrow 1$, $\mathcal{Z}$ is large
and its argument depend on the sign of $\eta -1$. Using appropriate
asymptotic approximations of WPCF with large argument, it can be shown that%
\begin{equation}
\Psi \left( a,c;z\right) \approx \left( \eta -1\right) ^{-a}c^{-a}\,,\ \
\left\vert \eta -1\right\vert \rightarrow 1\,,  \label{ap8}
\end{equation}%
if $\eta -1>0$ and%
\begin{equation}
\Psi \left( a,c;z\right) \approx \left( 1-\eta \right) ^{-a}c^{-a}\left\{ 
\begin{array}{l}
e^{i\pi a}\,,\ 0\leq \arg c<\pi \,, \\ 
e^{-i\pi a}\,,\ -\pi \leq \arg c<0\,,%
\end{array}%
\right.  \label{ap10}
\end{equation}%
as $\left\vert \eta -1\right\vert \rightarrow 1$ if $\eta -1<0$, both in
leading-order approximation. In Eq. (\ref{ap10}) note that $\arg \mathcal{Z}%
=-\pi +\left( \arg c\right) /2$\ if\ $0\leq \arg c<\pi $ and $\arg \mathcal{Z%
}=\pi +\left( \arg c\right) /2$, if $-\pi \leq \arg c<0$.

For large $\mu \rightarrow \infty $ and bounded $\left\vert z\right\vert $, $%
\left\vert \kappa \right\vert $, the asymptotic approximation \cite{Buchholz}
\begin{equation}
M_{\kappa ,\mu }\left( z\right) \approx z^{\mu +\frac{1}{2}}\,,\left\vert
\arg \left( \mu \right) \right\vert \leq \pi /2\,,  \label{ap11}
\end{equation}%
and the connection formulae%
\begin{eqnarray}
W_{\kappa ,\mu }\left( z\right) &=&\frac{\pi }{\sin 2\pi \mu }\left\{ -\frac{%
M_{\kappa ,\mu }\left( z\right) }{\Gamma \left( \frac{1}{2}-\mu -\kappa
\right) \Gamma \left( 1+2\mu \right) }\right.  \nonumber \\
&+&\left. \frac{M_{\kappa _{2},-\mu _{2}}\left( z_{2}\right) }{\Gamma \left( 
\frac{1}{2}+\mu _{2}-\kappa _{2}\right) \Gamma \left( 1-2\mu _{2}\right) }%
\right\} \,,  \nonumber \\
W_{-\kappa ,\mu }\left( e^{\pm i\pi }z\right) &=&\frac{\pi }{\sin 2\pi \mu }%
\left\{ \frac{\exp \left[ \pm i\pi \left( -\mu +1/2\right) \right] }{\Gamma
\left( \frac{1}{2}+\mu +\kappa \right) }\frac{M_{\kappa ,-\mu }\left(
z\right) }{\Gamma \left( 1-2\mu \right) }\right.  \nonumber \\
&-&\left. \frac{\exp \left[ \pm i\pi \left( \mu +1/2\right) \right] }{\Gamma
\left( \frac{1}{2}-\mu +\kappa \right) }\frac{M_{\kappa ,\mu }\left(
z\right) }{\Gamma \left( 1+2\mu \right) }\right\} \,,  \label{ap2.2}
\end{eqnarray}%
can be used to derive a asymptotic approximations for $W_{-\kappa _{1},\mu
_{1}}\left( e^{-i\pi }z_{1}\right) $ and $W_{\kappa _{2},\mu _{2}}\left(
z_{2}\right) $. Setting $\mu =\mu _{1}$ and $\kappa =\kappa _{1}$, both
defined in Eqs. (\ref{21.4}), we select $\chi =-1$, to find%
\begin{eqnarray}
&&\left. W_{-\kappa _{1},\mu _{1}}\left( e^{-i\pi }z_{1}\right) \right\vert
_{t=0}\approx \frac{e^{-\frac{i\pi }{4}}}{\sqrt{\sinh \left( 2\pi eE\tau
_{1}^{2}\right) }}  \label{ap2.3} \\
&&\times \left[ \frac{e^{i\Theta _{1}^{-}}e^{-\frac{i\pi \mu _{1}}{2}}}{%
i\tau _{1}}\sqrt{\frac{\lambda \sinh \left( \pi \nu _{1}^{+}\right) }{eE}}%
+e^{i\Theta _{1}^{+}}e^{\frac{i\pi \mu _{1}}{2}}\sqrt{\sinh \left( \pi \nu
_{1}^{-}\right) }\right] \,,  \nonumber
\end{eqnarray}%
as $\left\vert \mu _{1}\right\vert \rightarrow \infty $ for Fermions in
next-to-leading order approximation. As for the Whittaker function $%
W_{\kappa _{2},\mu _{2}}\left( z_{2}\right) $, one finds%
\begin{eqnarray}
W_{\kappa _{2},\mu _{2}}\left( z_{2}\right) &\approx &e^{i\Theta
_{2}^{+}}\exp \left( -\frac{\pi eE\tau _{2}^{2}}{2}\right) \sqrt{2}\sqrt{%
\frac{\sinh \pi \nu _{2}^{-}}{\sinh \left( 2\pi eE\tau _{2}^{2}\right) }} 
\nonumber \\
&+&e^{i\Theta _{2}^{-}}\exp \left( \frac{\pi eE\tau _{2}^{2}}{2}\right) 
\frac{\sqrt{\lambda }}{\sqrt{2eE}\tau _{2}}\sqrt{\frac{\sinh \pi \nu _{2}^{+}%
}{\sinh \left( 2\pi eE\tau _{2}^{2}\right) }}\,,  \label{ap2.4}
\end{eqnarray}%
as $\left\vert \mu _{2}\right\vert \rightarrow \infty $ for Fermions in
next-to-leading order approximation. Similar expansions can be obtained for
the Klein-Gordon case. The complex phases in both equations are $\Theta
_{j}^{\pm }=-\arg \Gamma \left( \mp i\nu _{j}^{\mp }\right) -\arg \Gamma
\left( \pm 2ieE\tau _{j}^{2}\right) \pm eE\tau _{j}^{2}\ln \left( 2\omega
_{j}\tau _{j}\right) $.

For small $z$, bounded $\kappa $ and $\mu =1/2$, the Whittaker function
acquires the series expansion \cite{DLMF}%
\begin{equation}
W_{\kappa ,\frac{1}{2}}\left( z\right) =\frac{1}{\Gamma \left( 1-\kappa
\right) }+\frac{1}{2\Gamma \left( -\kappa \right) }\left\{ \frac{1}{\kappa }%
+2\left[ -1+2\gamma +\log \left( z\right) +\psi \left( 1-\kappa \right) %
\right] z+O\left( z^{2}\right) \right\} \,,\ \ z\rightarrow 0\,,
\label{ap2.5}
\end{equation}%
where $\gamma \approx 0.577$ is Euler's constant $\psi \left( z\right)
=\Gamma ^{\prime }\left( z\right) /\Gamma \left( z\right) $ is the Psi (or
DiGamma) function.

\end{document}